%% file: main.tex
\documentclass[reprint, 
 amsmath,amssymb,
 aps, physrev,
pra,
twocolumn,
superscriptaddress
]{revtex4-2}
\usepackage[utf8]{inputenc}
\usepackage{graphicx}
\usepackage{gensymb}
\usepackage{rotating}
\usepackage{upgreek}
\usepackage{xcolor}

\newcommand{\overbar}[1]{\mkern 1.5mu\overline{\mkern-1.5mu#1\mkern-1.5mu}\mkern 1.5mu}

\begin{document}


\title{\textbf{Bias-field-free operation of nitrogen-vacancy ensembles in diamond for accurate vector magnetometry}
}%
\author{Lilian Childress}
\email{lilian.childress@mcgill.ca}
\affiliation{Department of Physics, McGill University, 3600 Rue University, Montreal Quebec H3A 2T8, Canada}

\author{Vincent Halde}
\email{vincent.halde@sbquantum.com}
\affiliation{
 SBQuantum, 805 Rue Galt O, Sherbrooke, Quebec J1H 1Z1, Canada
}
\author{Kayla Johnson}
\affiliation{
 SBQuantum, 805 Rue Galt O, Sherbrooke, Quebec J1H 1Z1, Canada
}
\author{Andrew Lowther}
\affiliation{
 SBQuantum, 805 Rue Galt O, Sherbrooke, Quebec J1H 1Z1, Canada
}
\author{David Roy-Guay}
\affiliation{
 SBQuantum, 805 Rue Galt O, Sherbrooke, Quebec J1H 1Z1, Canada
}
\author{Romain Ruhlmann}
\affiliation{
 SBQuantum, 805 Rue Galt O, Sherbrooke, Quebec J1H 1Z1, Canada
}
\author{Adrian Solyom}
\affiliation{
 SBQuantum, 805 Rue Galt O, Sherbrooke, Quebec J1H 1Z1, Canada
}

\date{\today}

\begin{abstract}
Accurate measurement of vector magnetic fields is critical for applications including navigation, geoscience, and space exploration. Nitrogen-vacancy (NV) center spin ensembles offer a promising solution for high-sensitivity vector magnetometry, as their different orientations in the diamond lattice measure different components of the magnetic field. However, the bias magnetic field typically used to separate signals from each NV orientation introduces inaccuracy from drifts in permanent magnets or coils. Here, we present a novel bias-field-free approach that labels the NV orientations via the direction of the microwave (MW) field in a variable-pulse-duration Ramsey sequence used to manipulate the spin ensemble. Numerical simulations demonstrate the possibility to isolate each orientation's signal with sub-nT accuracy in most terrestrial fields, even without precise MW field calibration, at only a moderate cost to sensitivity. We also provide proof-of-principle experimental validation, observing relevant features that evolve as expected with applied magnetic field. Looking forward, by removing a key source of drift, the proposed protocol lays the groundwork for future deployment of NV magnetometers in high-accuracy or long-duration missions. 

\end{abstract}

\maketitle


\section{\label{sec:level1}Introduction}
Magnetic sensor development often focuses on precision, the ability to sense as small a change as possible. Nevertheless, in some contexts accuracy or long-term stability can prove more important. Accurate maps of the earth's magnetic field are critical to navigation~\cite{canciani_airborne_2017, gupta_lower_2024, canciani_magnetic_2022, huang_tightly-integrated_2022}, geoscience~\cite{alken_international_2021, olson_changes_2006, balasis_investigation_2023, heavlin_case-control_2022}, and resource exploration~\cite{nabighian_75th_2005, clark_new_2012, stolz_squids_2022}; sensors that retain calibration over months or years are essential to exploratory space missions~\cite{bennett_precision_2021}. 
While scalar atomic magnetometers can exhibit absolute accuracy down to $\sim$50 pT~\cite{jager_swarm_2010, noauthor_gem_nodate}, vector mode operation remains a challenge, with field-deployed vector-mode atomic magnetometers reaching only $\sim$nT accuracy~\cite{hulot_swarms_2015, fratter_swarm_2016}. Even hybrid solutions employing a scalar atomic magnetometer for {\it in-situ} calibration of a vector fluxgate sensor attain only slightly better accuracy, with the additional cost of complex calibration procedures and reduced performance in high field gradients~\cite{toffner-clausen_-flight_2016, gravrand_calibration_2001}. 

\begin{figure}[ht]
    \includegraphics{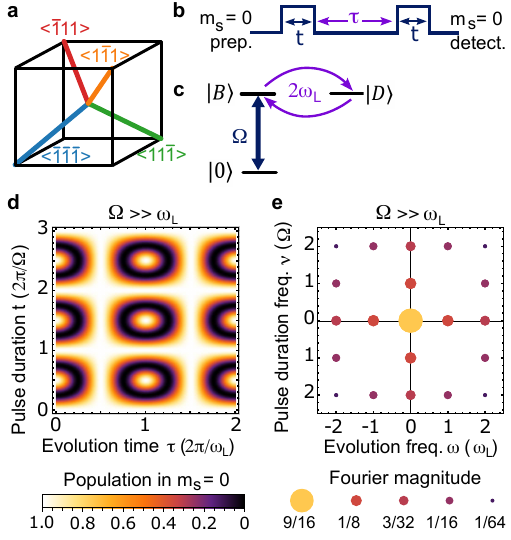}
    \caption{(a) Illustration of the four possible directions of the NV symmetry axis (N-to-V axis) in the conventional diamond crystal unit cell. (b) Variable-pulse-duration Ramsey (VPDR) sequence with two microwave (MW) pulses of duration $t$ separated by free evolution time $\tau$. (c) Illustration of dynamics in the bright-dark basis. 
    (d) Population in $|0\rangle$ ($P_0(t,\tau)$) after application of the pulse sequence in (b) in the $\Omega/\omega_L \rightarrow \infty$ limit, with constant MW phase, no detuning, no dephasing, and an axial magnetic field. 
    The pulse duration $t$ is given in units of the Rabi period and the evolution time $\tau$ is given in units of the Larmor period. (e) Magnitude of the 2D Fourier transform of the population in (d), with discrete frequencies illustrated with dots of area proportional to their magnitude. The quantitative values given in the legend are $|a_{nm}|$ in the Fourier expansion $P_0(t,\tau) = \sum_n \sum_m a_{nm}e^{i \nu_n t}e^{i\omega_m \tau}$. The frequency associated with the pulse duration (free evolution time) is given in units of the Rabi frequency $\Omega$ 
    (Larmor frequency $\omega_L$).}
    \label{fig1}
\end{figure}

Nitrogen-vacancy (NV) centers in diamond offer an appealing alternative for accurate vector magnetometry because they combine the long coherence time and Hamiltonian-defined response of atomic magnetometers with intrinsic vector measurement capabilities arising from their site symmetry~\cite{rondin_magnetometry_2014, barry_sensitivity_2020}. Each of the four possible orientations for the NV in the diamond lattice (see Fig.~\ref{fig1}a) is most sensitive to fields along its symmetry axis, permitting vector field measurement. To reconstruct the magnetic field, one must know which signals come from which NV orientation.  The most common solution applies a bias field of a few mT using permanent magnets or coils to separate the spin transition frequencies of different orientations, allowing vector measurement of small shifts due to the external magnetic field~\cite{maertz_vector_2010, steinert_high_2010}. This approach has enabled NV magnetometers with sufficient accuracy for selection as finalist in the MagQuest Challenge~\cite{oshea_sbquantums_2023, halde_who_2025}, signaling the potential for NVs to outclass existing magnetometers for earth field mapping. However, at the nT accuracy level the bias field becomes problematic because it is indistinguishable from the external field, such that drifts in bias due to thermal~\cite{danieli_highly_2013}, mechanical, or hysteresis~\cite{moree_review_2023} effects induce systematic errors. Removing large bias fields could also aid materials research applications, as $\sim$ mT fields are incompatible with some target samples~\cite{wang_zero-field_2022}. 
Bias-field-free operation is thus desirable to enable high-accuracy NV vector magnetometry and expand its range of applications. 

Indeed, there has been a recent surge of interest in bias-field-free schemes. Several groups have explored the use of optical anisotropy, labeling orientations via distinct spatial emission patterns~\cite{backlund_diamond-based_2017, chen_calibration-free_2020, weggler_determination_2020} 
or excitation and emission polarization~\cite{reuschel_vector_2022,munzhuber_polarization-assisted_2020, li_simultaneous_2024, li_vector_2024}. When applied to NV ensembles, however, these schemes incompletely isolate the contributions from different orientations, such that any drifts in the optical calibration will degrade accuracy, particularly when two orientations experience similar magnetic field projections such that their spin transitions nearly coincide and relevant features overlap. Microwave (MW) polarization has also been suggested to discern vector fields from a single orientation~\cite{isogawa_vector_2023} or label orientations, in the latter case either on its own~\cite{yoon_identifying_2025} or combined with optical polarization ~\cite{munzhuber_polarization-assisted_2020}. These proposed implementations using MW polarization exhibit accuracy issues similar to those of optical polarization schemes. While circular MW polarization in combination with optimal control could in principle provide better isolation of each orientation's signal~\cite{liddy_optimal_2023}, robust implementation over a terrestrial-magnetic-field range remains to be demonstrated. Pulsed biasing~\cite{pham_chopped_2021} could improve accuracy if a sufficiently precise control system can be realized, but it would not function for samples requiring low field. Thus it remains an open challenge to identify a robust, cross-talk-free, drift-free mechanism for labeling NV orientations in the absence of a bias field.

Here, we examine a novel method to label NV orientations via the MW field. By appropriately adjusting the MW field angle relative to the diamond crystalline axes, each NV orientation experiences a distinct Rabi frequency. 
While Rabi-labeling has been proposed before~\cite{roy-guay_vectorial_2021}, we identify a technique that permits effective use of the Rabi-frequency label without relying on knowledge of the exact MW direction or a calibrated model of the NV spin resonance lineshapes. 
In particular, we employ a Ramsey-like protocol with MW pulses of variable duration (see Fig.~\ref{fig1}b), separating the labeling dimension (pulse duration) from the sensing dimension (interpulse delay) to enable true isolation of signals originating from each orientation. Crucially, the variable-pulse-duration Ramsey (VPDR) protocol allows us to identify drifts in the MW field separately from the target DC field, such that precise calibration of the MW magnitude and direction is not required. 
In addition, the VPDR sequence probes double quantum spin transitions, which enhances stability in the presence of temperature fluctuations and eliminates dephasing from axial strain inhomogeneities~\cite{barry_sensitivity_2020}. 
The scheme requires only a single MW frequency, easing experimental implementation; while it cannot operate at true zero field, for reasonable parameters it has a range exceeding requirements for sensing in terrestrial fields.   

The paper is structured as follows. In section~\ref{sec:theory} we analytically model the response of a single NV to the VPDR pulse sequence and analyze it in the frequency domain. This model provides insight into different regimes of operation and allows estimation of the sensitivity cost relative to more standard measurements. We then test the VPDR protocol on numerical simulations of an ensemble of NV spins in section~\ref{sec:numerics}, with open-source code for the simulator provided~\cite{noauthor_httpsgithubcomsbquantum2sbq-dq-simulator_2025}. We demonstrate that it is possible to suppress cross-talk between signals from different orientations to reach sub-nT accuracy across an earth-magnetic-field range, apart from discrete dead zones that we identify and discuss. Moreover, we show that similar accuracy is retained even for factor-of-two changes in MW magnitude and $\pm 4\degree$ MW direction fluctuations, such that accuracy is not constrained by MW calibration precision. Finally, in section~\ref{sec:experiment} we show a proof-of-principle experimental demonstration, comparing observed features to those expected from the analytic theory, illustrating the feasibility of Rabi-labeled, bias-field-free vector field measurements. 

\section{Theoretical principles \label{sec:theory}}

\subsection{Single NV response}
The VPDR pulse sequence 
(see Fig.~\ref{fig1}b) comprises two microwave (MW) pulses of duration $t$ separated by a free evolution time $\tau$, where the second MW pulse 
has a phase shift $\phi$ of $0\degree$ or $180\degree$ with respect to the first. 
We begin by analyzing the effect of the VPDR sequence on a single NV center modeled as a spin $S = 1$ system with a zero-field splitting (ZFS) $\Delta/(2\pi)$ that we set to  2.87 GHz~\cite{cambria_physically_2023}. We assume that the NV spin is initially polarized into the $m_s = 0$ Zeeman projection ($|0\rangle$), and that a small axial field $B_z$ (to be measured) 
splits the $m_s = \pm 1$ states ($|\pm1\rangle$) by energy $2\hbar \omega_L = 2g \mu_B B_z$. A MW field $\mathbf{B_\text{MW}} \cos{(\nu_{MW} t + \phi)}$ is applied to the spin. By choosing the $\hat{z}$ coordinate axis along the NV symmetry axis and $\hat{x}$ along the projection of the MW field in the plane perpendicular to $\hat{z}$ ($\mathbf{B_{MW, \perp}}$), the NV spin Hamiltonian can be written in the rotating frame and rotating wave approximation as
\begin{eqnarray}
    H &=& -\hbar \delta \hat{S}_z^2 + \hbar \omega_L \hat{S}_z \nonumber \\&&+ \frac{\hbar\Omega}{2} \left(\cos{\phi}\hat{S}_x + \sin{\phi}\left(\hat{S}_y\hat{S}_z + \hat{S}_z\hat{S}_y\right)\right) 
\label{eq:H}
\end{eqnarray}
where $\Omega = g \mu_B B_{MW, \perp}/\hbar$ is the Rabi frequency, $\delta =  \nu_{MW} - \Delta$ is the MW detuning from the ZFS (unless otherwise specified we will assume $\delta = 0$), and $\hat{S}_{x,y,z}$ are the spin-1 matrices. Note that the MW phase $\phi$ is not analogous to the MW direction for a spin-1 system.

\subsubsection{The high Rabi frequency limit}

By propagating a $|0\rangle$-initialized spin through the three intervals of the pulse sequence 
we can obtain an expression for the final $m_s = 0$ population $P_0(t,\tau)$ (see Appendix~\ref{app:analytic}). 
In the hard-pulse limit $\Omega/\omega_L\rightarrow \infty$ and $\Omega/\delta \rightarrow \infty$, 
\begin{eqnarray}
    \lim_{\substack{\Omega/\omega_L \rightarrow \infty\\ \Omega/\delta \rightarrow \infty}} 
    P_0(t, \tau) &\approx& \cos^4{\frac{\Omega t}{2}} \nonumber\\&&+ \underbrace{\frac{\cos{(\tau \delta + \phi)}}{2}\sin^2{\Omega t}\cos{\omega_L \tau}}_{SQ}\nonumber\\&&+ \underbrace{\sin^4{\frac{\Omega t}{2}}\cos^2{\omega_L \tau}}_{DQ}.
\label{eq:P0}
\end{eqnarray}
An illustration of this signal for $\delta = 0, \phi = 0$ is shown in Fig.~\ref{fig1}d. Examining Eq.~\ref{eq:P0}, two of its terms exhibit oscillations with $\tau$ that can be used to deduce the Larmor precession frequency (and thus the magnetic field). The second term oscillates at $\omega_L$ (a single-quantum ($SQ$) Ramsey signal), while the third term oscillates at $2\omega_L$ (a double-quantum ($DQ$) Ramsey signal). 

Only single quantum terms are sensitive to the detuning or phase of the microwaves. In particular, changing the phase of the second pulse by $180\degree$ reverses the sign of the $SQ$ term, similar to techniques employed for double-quantum Ramsey in a bias field~\cite{hart_nv-diamond_2021}. Thus, by applying the pulse sequence twice, once with $\phi = 0\degree$ and once with $\phi = 180\degree$, and adding the resulting signals, the $SQ$ contribution can be canceled. 
Isolating the $DQ$ signal is beneficial because of its insensitivity to detuning (and thus to variations in temperature and axial strain); moreover, $DQ$ transitions do not suffer from dephasing due to axial strain inhomogeneities, offering an opportunity for enhanced magnetic sensitivity~\cite{fang_high-sensitivity_2013, mamin_multipulse_2014, moussa_preparing_2014, bauch_ultralong_2018}.

Because $P_0(t,\tau)$ is a sum of oscillations, it is natural to examine its Fourier transform $\mathcal{P}_0(\nu, \omega) = \int dt \int d\tau e^{-i\nu t}e^{-i\omega \tau}P_0(t,\tau)$. In the absence of dephasing, $\mathcal{P}_0(\nu, \omega)$ is a sum of delta functions; Figure~\ref{fig1}e shows $\mathcal{P}_0(\nu, \omega)$ in the $\Omega /\omega_L \rightarrow \infty$ limit, with the magnitude of the area under each 2D delta function 
given by the area of the circle centered at its frequencies $(\nu,\omega)$.  
This Fourier domain representation enables a simple conceptual interpretation of data: the vertical location of a Fourier peak shows the Rabi frequency (which will be used to label orientation) while its horizontal location indicates the free precession frequency associated with that Rabi frequency, thus revealing information about the projection of the magnetic field on that Rabi-labeled NV axis.

For quantitative field estimation, we will focus on the $DQ$ term at $\nu = \Omega, \omega = 2\omega_L$. 
The origin of this term can be 
understood in the bright/dark state picture (see Fig.~\ref{fig1}c). The pulsed MW drive couples $|0\rangle$ to the so-called bright state $|B\rangle = \frac{1}{\sqrt{2}}\left(|+1\rangle + |-1\rangle\right)$, producing oscillations between $|0\rangle$ and $|B\rangle$ at the Rabi frequency $\Omega$. 
During free evolution, the axial magnetic field induces oscillations between $|B\rangle$ and the orthogonal dark state $|D\rangle = \frac{1}{\sqrt{2}}\left(|+1\rangle - |-1\rangle\right)$ 
at twice the Larmor precession frequency $2\omega_L$. 
Whenever the pulse duration $t$ is an odd integer multiple of $\pi/\Omega$, each pulse will transfer all the population into or out of the $\{|B\rangle, |D\rangle\}$ subspace, leading to a final $|0\rangle$ population that oscillates strongly with $\tau$
; whenever $t$ is an even multiple of $\pi/\Omega$, the $\{|B\rangle, |D\rangle\}$ subspace will not be populated during the free evolution period and the Larmor precession signal vs $\tau$ will be suppressed. 
We thus expect the final $m_s = 0$ population to contain an oscillation with $\tau$ at 
frequency $2\omega_L$ modulated by an oscillation with $t$ at the Rabi frequency $\Omega$. 


It is straightforward to extract this $DQ$ term of interest from a data set 
acquired at a range of pulse durations $t$ and one or more free evolution times $\tau$. By taking the inner product of $P_0(t,\tau)$ with $\cos{\Omega t}$ along the pulse-duration dimension, one obtains the double-quantum Ramsey signal 
\begin{equation}
\lim_{\substack{\Omega/\omega_L \rightarrow \infty\\ \Omega/\delta \rightarrow \infty\\T\rightarrow\infty}}\frac{1}{T}\int_{0}^{T} 2\cos{\Omega t}~P_0(t, \tau)dt = \frac{1}{4}\left(1-\cos{2\omega_L \tau}\right),
\label{eq:innerP0}
\end{equation}
from which the magnetic field along the NV axis can be inferred. 

The theoretical sensitivity of the VPDR protocol is negatively impacted by taking measurements at pulse durations $t$ that are not optimal. Nevertheless, the cost is not overwhelming: in the hard-pulse limit, for data analyzed via a cosine inner product on the pulse-duration dimension, sampling at many pulse durations only worsens the sensitivity by a factor of $2\sqrt{2}$ relative to a more standard double-quantum Ramsey experiment employing $t = \pi/\Omega$ . While this result holds even in the presence of dephasing during free evolution, it is valid only in the high Rabi frequency limit. For moderate Rabi frequencies, dephasing during MW pulses can further degrade VPDR sensitivity (see Appendix~\ref{app:sensitivity} for details).

\begin{figure}[ht]
    \includegraphics{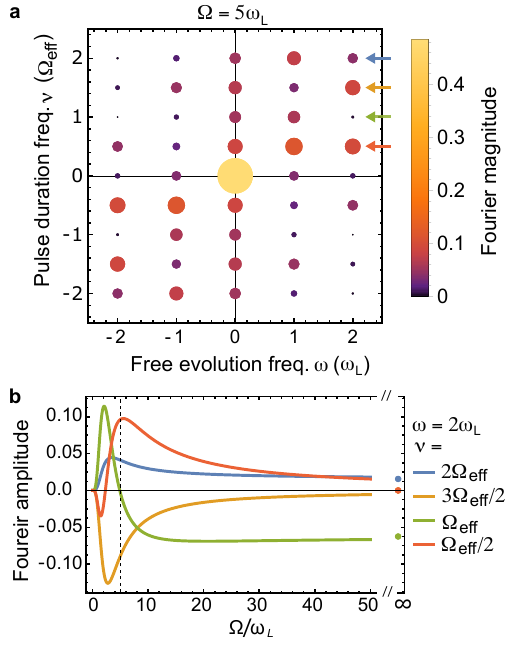}
    \caption{(a) $\mathcal{P}_0(\nu, \omega)$ for $\Omega = 5 \omega_L$, no detuning or dephasing, and constant MW phase. The area of the dots is proportional to the magnitude of the discrete Fourier components as in Fig.~\ref{fig1}e. The pulse duration frequency $\nu$ is given in units of the effective Rabi frequency $\Omega_{\mathrm{eff}} = \sqrt{\Omega^2 + 4\omega_L^2}$ while the free evolution frequency is given in units of the Larmor frequency $\omega_L$. (b) Amplitude of the positive-quadrant $DQ$ Fourier components, 
    corresponding to the four components indicated with arrows in (a), as a function of $\Omega/\omega_L$. The dotted line indicates the value of $\Omega/\omega_L$ for which (a) is evaluated.}
    \label{fig2}
\end{figure}

\subsubsection{Finite $\Omega/\omega_L$ considerations}

The simple expression of Eq.~\ref{eq:P0} is valid only in the $\Omega/\omega_L \rightarrow \infty,~ \Omega/\delta \rightarrow \infty$ limit. While we can impose $\delta = 0$, it is important to understand the effect of finite $\Omega/\omega_L$. The Larmor precession of the spin during the pulse modifies the response of the system in two ways: First, the $|0\rangle \rightarrow |B\rangle$ population oscillations occur at the effective Rabi frequency $\Omega_\text{eff} = \sqrt{\Omega^2 + 4 \omega_L^2}$; secondly, new oscillations with $t$ emerge at $\nu = \pm \Omega_\text{eff}/2$ and  $\nu = \pm 3\Omega_\text{eff}/2$. The full analytic expressions for $\delta = 0$ are given  in Appendix~\ref{app:analytic}, and the resulting signals can be understood conceptually in the Fourier domain as illustrated in Fig.~\ref{fig2}a. 
Note that $\nu$ is given in units of $\Omega_\text{eff}$ (not $\Omega$), while $\omega$ is given in units  of $\omega_L$. Figure~\ref{fig2}a clearly shows the emergence of new Fourier peaks, as well as changes to the magnitudes of the original peaks seen in Fig.~\ref{fig1}e. 

Perhaps surprisingly, the new Fourier components persist to fairly large values of $\Omega/\omega_L$. Figure~\ref{fig2}b shows the Fourier amplitudes of the $DQ$ terms in the first quadrant, as indicated by arrows in Fig.~\ref{fig2}a, as a function of $\Omega/\omega_L$.  In particular, the $\nu = \Omega_\text{eff}/2$ term remains appreciable even at $\Omega/\omega_L = 50$. In contrast, the term of interest at $\nu = \Omega_\text{eff}$ (green line) approaches its saturating value for $\Omega/\omega_L \gtrsim 10$. This behavior implies that analysis will be simpler 
when $\Omega$ exceeds $\omega_L$ by an order of magnitude or more. 

Despite the additional signal complexity at finite $\Omega/\omega_L$, it remains possible to cancel the $SQ$ peaks at $\omega = \omega_L$ by adding $0\degree$  and $180\degree$-phase-shifted signals (see Appendix~\ref{app:analytic}). Since the $DQ$ signals are unaffected by MW phase, this $SQ$ cancellation procedure does not impact sensitivity, but implies a refresh rate reduction by a factor of 2.


    
\subsubsection{Dephasing}
Spin dephasing does not substantially change the interpretation of VPDR signals. In an analytic approximation valid for $\Omega \gg \omega_L$, we neglect spin dephasing during the short pulses and focus on dephasing during the longer free evolution period. Modeling the environmental noise as Lorentzian-distributed inhomogeneous variations in magnetic field~\cite{dobrovitski_decoherence_2008} produces an exponential decay of all $SQ$ signals as $e^{-\tau/T_2^*}$ and all $DQ$ signals as $e^{-2\tau/T_2^*}$ (see Appendix~\ref{app:analytic} for details). In the high-Rabi-frequency approximation of Eq.~\ref{eq:P0}, an inner product of the dephased signal with $\cos(\Omega t)$  yields  
$1/T\int_0^T 2\cos{\Omega t}~P_0(t, \tau)dt \approx \frac{1}{4}\left(1-e^{-2\tau/T_2^*}\cos{2\omega_L \tau}\right)$, which is half of the signal from a standard double-quantum Ramsey experiment, with the same decay term. As long as the dephasing during the pulses is negligible, a similar post-inner-product result can be obtained even at finite $\Omega/\omega_L$, however $SQ$ cancellation will be necessary, and the magnitude of the $DQ$ signal will vary with $\Omega/\omega_L$. Dephasing during the MW pulses does not lend itself to simple analytic treatment; as shown numerically in Appendix~\ref{app:sensitivity}, it further reduces the amplitude of the post-inner-product signal, worsening sensitivity but not impacting the functional dependence on $\tau$ (nor the resulting accuracy). 

\subsubsection{Hyperfine structure}

Finally, NV centers have hyperfine structure associated with the host nitrogen nuclear spin. We consider the dominant $I = 1$ isotope $^{14}$N as the $I = \frac{1}{2}$ $^{15}$N isotope would introduce additional modulation due to its lack of quadrupole splitting~\cite{oon_ramsey_2022}. In the secular approximation, the $^{14}$N nuclear spin adds an axial magnetic field proportional to its projection on the NV symmetry axis. When the nuclear spin is in Zeeman sublevel $m_I$, the effective Larmor frequency is $\omega_L = \omega_L^{ext} + m_I A$, where $\omega_L^{ext}$ is the Larmor precession frequency associated with an external axial magnetic field, and $A/(2\pi) = 2.16$ MHz~\cite{smeltzer_robust_2009}. 
Since the nuclear spin sublevels are approximately equally populated at room temperature and low magnetic field, we can account for hyperfine structure by averaging $P_0(t, \tau)$ over the three possible values of $m_I$. 
Qualitatively, hyperfine structure leads to a tripling of the Fourier-domain peaks in $\mathcal{P}(\nu, \omega)$ along the horizontal $\omega$ axis, corresponding to $\omega = m (\omega_L^{ext} + m_I A)$ with $m = \{-2, -1, 0, 1, 2\}$ and $m_I = \{-1, 0, 1\}$. With this additional complication, it becomes especially beneficial to cancel the $SQ$ contribution by adding a $180 \degree$ phase-shifted signal, such that only $m = \{-2, 0,2\}$ Fourier components are present in the signal. For the remaining theoretical and numerical analysis, we will work with such $SQ$-cancelled signals. 

 Hyperfine structure also induces variation of the effective Rabi frequency with nuclear spin projection, $\Omega_\text{eff}(m_I)= \sqrt{\Omega^2 + 4(\omega_L^{ext} + m_I A)^2}$. Extracting all three hyperfine signals from an inner product with a single cosine requires $\Omega \gg A$, further motivating the use of high Rabi frequency in an experimental realization. 

\subsection{Ensemble signals}

\begin{figure}[ht!]
    \includegraphics{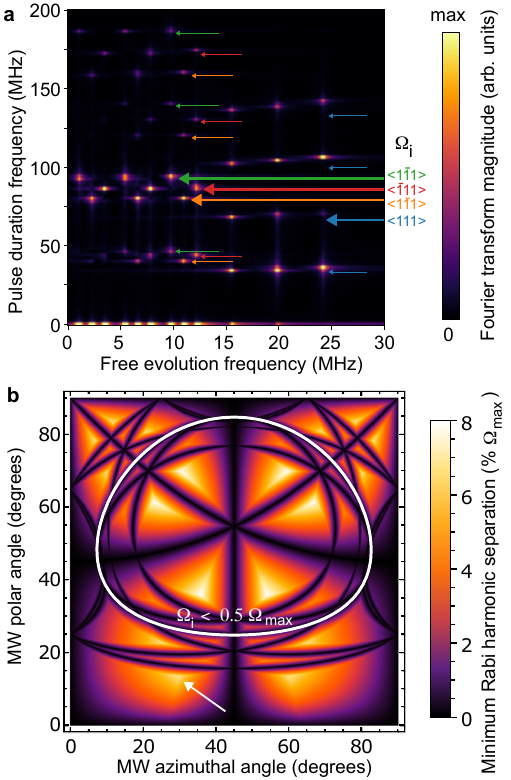}
    \caption{(a) The discrete Fourier transform (positive-frequency quadrant only) of a simulated VPDR Ramsey signal with SQ cancellation for an NV ensemble. $\Omega_\text{max}/2\pi = $ 100 MHz, $T_2^* = 2~\upmu$s, $\mathrm{B} = (185, 204, 223)~\upmu$T in the conventional diamond crystal coordinates, and $\mathbf{B_\text{MW}} \parallel (0.2054, .1188, .9714)$ corresponding to the locally optimal orientation indicated in (b). Thick arrow-tipped lines indicate each orientation's bare Rabi frequency $\Omega_i$, 
    while thin arrow-tipped lines indicate $\Omega_i/2, 3\Omega_i /2$ and $2\Omega_i $. (b) Optimization of the MW magnetic field direction. The minimum separation between $\Omega_i$ and $n\Omega_j$ for NV orientations $i\neq j$ and $n \in \{\frac{1}{2},1,\frac{3}{2},2\}$ is given 
    as a function of the direction of the MW magnetic field $\mathbf{B_\text{MW}}$ relative to the conventional diamond unit cell. 
    For directions inside the white contour, one of the $\Omega_i$ is less than $\Omega_\text{max}/2$; the white arrow indicates an optimal orientation outside the contour.   }
    \label{fig3}
\end{figure}

When all four orientations of NV centers are present, each species $i$ will have its own Rabi frequency $\Omega_i = (g\mu_B/\hbar) | \mathbf{B_\text{MW}}\times \hat{z}_i|$ set by the perpendicular projection of the MW magnetic field $\mathbf{B_\text{MW}}$ on the $i^{th}$ symmetry axis $\hat{z}_i$. Figure~\ref{fig3}a shows an example of the Fourier transform of a numerically calculated ensemble signal (see section~\ref{sec:numerics}) with SQ cancellation for a magnetic field  $|B| \approx 0.35$~mT slightly canted from the $\langle 111\rangle$ direction. 
As expected, the signals coming from the $\langle 111\rangle$ subensemble 
occur at significantly higher free evolution frequency than signals from the other three orientations. 
Within the first quadrant, each orientation $i$ exhibits 12 $DQ$ Fourier components occurring at $\nu = n\Omega_{\text{eff},i}$ for $n \in \{\frac{1}{2}, 1, \frac{3}{2}, 2\}$ and $\omega = 2|\omega_{L,i}^{ext} + m_I A|$ for $m_I \in \{-1, 0, 1\}$.  
The bare Rabi frequencies $\Omega_i$ (as well as their harmonics) are illustrated with arrow-tipped horizontal lines, identifying 
each orientation.

A few features of Fig.~\ref{fig3}a are worth noting.  Owing to the large $B_{DC}$ value, $\Omega_{\text{eff},i}$ differs from $\Omega_i$, especially for the $\langle 111\rangle$ orientation. The relatively small value of $\Omega_i/(\omega_{L,i}^{ext} + m_I A)$ is also responsible for the weak magnitude of the $\langle111\rangle$ peaks at $\nu = \Omega_{\text{eff},\langle 111\rangle}$. 
Both of these effects would be strongly suppressed for smaller $B_{DC}$ within the range of earth's field. More importantly, Fig.~\ref{fig3}a demonstrates an advantage of the VPDR protocol over previous polarization-based approaches to bias-field-free orientation labeling~\cite{backlund_diamond-based_2017, reuschel_vector_2022, munzhuber_polarization-assisted_2020}. For a field direction near $\langle 111\rangle$, the three off-axis orientations have nearly commensurate spin transition frequencies, such that their ODMR features would overlap. By separating the orientations' signals along the pulse duration dimension, the VPDR protocol ensures that their transition frequencies can be independently measured. 

To fully isolate each orientation's signal, the Rabi frequencies $\Omega_i$ should be well separated, such that, for example, taking an inner product of the signal with $\cos{\Omega_i t}$ picks out the $DQ$ signal from orientation $i$ with minimal crosstalk from other orientations. Qualitatively, the best results will be obtained when the 
orientations' Rabi frequencies and their harmonics (arrows in 
in  Fig.~\ref{fig3}a) do not come too close to each other, which can be controlled with the MW direction. 
We optimize the MW direction by calculating the smallest separation $|\Omega_i - n\Omega_j|$ for $i\neq j$ 
with $n \in \{\frac{1}{2}, 1, \frac{3}{2}, 2\}$ while varying the polar and azimuthal angles of $\mathbf{B_\text{MW}}$ relative to the conventional diamond unit cell. Figure~\ref{fig3}b shows a plot of this smallest separation expressed as a percentage of the maximum Rabi frequency $\Omega_\text{max} = (g\mu_B/\hbar) |B_{MW}|$. Minimum separations just over $8\%$ of $\Omega_\text{max}$ are possible, but they are achieved by making one of the $\Omega_i$ much smaller than the others, which could be problematic for experiments with limited $\Omega_\text{max}$. The local maximum near $\theta_{MW} = 13.74\degree, \phi_{MW} = 30.05\degree$ with $6.8\%$ minimum separation (white arrow) may prove a better candidate because all four $\Omega_i$ exceed $0.65 \Omega_\text{max}$. For Fig.~\ref{fig3}a and the numerical analysis in section~\ref{sec:numerics}, we employ this MW orientation.

In an ensemble, off-axis DC magnetic fields are generally present for each NV orientation. The off-axis components of the field will partially hybridize the eigenstates of the free-evolution Hamiltonian and shift their eigenvalues. In particular, the evolution with $\tau$ will now occur at the frequency difference $\Delta \omega_i$ between the $m_s = \pm1$-like eigenstates of the $i^{th}$ orientation. Nevertheless, it is meaningful to extract the $\Delta \omega_i$ as they can be used to reconstruct the magnetic field via least-squares fitting~\cite{steinert_high_2010, garsi_three-dimensional_2024}, linearized Hamiltonian inversion~\cite{schloss_simultaneous_2018}, or 
machine learning techniques~\cite{zhang_deep-neural-network-based_2024}. Indeed, vector magnetometers based on continuous-wave optically detected magnetic resonance also measure eigenvalue differences, so when evaluating the performance of the VPDR protocol we will focus on accurately extracting the $\Delta\omega_i$ values. While the effects of off-axis fields are expected to be negligibly small in terrestrial fields (suppressed by $\sim(B_{\perp}/\Delta)^2$), we include them in the numerical simulations discussed below.


\section{\label{sec:numerics} Numerical simulations}
Our motivation for the VPDR protocol is the elimination of bias magnets for reduced drift and increased accuracy. It is thus essential to evaluate (1) the feasibility of accurately extracting $\Delta \omega_i$ from VPDR data without crosstalk from other orientations and (2) the robustness of this inversion to drifts in MW field amplitude or orientation. 
To these ends, we numerically simulate the response of an ensemble of NV centers. After projecting the external DC and MW magnetic fields onto each of the four possible NV coordinate systems, we solve the Master equation for each spin species (see Appendix~\ref{app:numerics} for calculation details), including off-axis magnetic fields, Markovian dephasing throughout the pulse sequence, and hyperfine interactions in the secular approximation; the code is publicly available~\cite{noauthor_httpsgithubcomsbquantum2sbq-dq-simulator_2025}. The simulation returns a $SQ$-cancelled signal $$S_\text{VPDR}(t,\tau) = P_0^\text{total}(t,\tau, \phi = 0) + P^\text{total}_0(t,\tau, \phi = \pi)$$ where $P^\text{total}_0(t,\tau, \phi)$ is the probability for an $m_s = 0$-initialized spin in the ensemble to end up in $|0\rangle$ after a VPDR pulse sequence with relative MW pulse phase $\phi$, averaged over NV orientation and nuclear spin projection. In particular, $S_\text{VPDR}(t,\tau)$ should be proportional to the sum of signals from fluorescence-based readout of the ensemble after the $\phi = 0$ and $\phi = \pi$ VPDR sequences. 

\subsection{\label{sec:inversion}Extracting transition frequencies for each orientation}
First, we examine whether we can accurately extract each orientation's $m_s = \pm 1$ transition frequency $\{\Delta \omega_i\}$ from a simulated data set $S_\text{VPDR}(t_j, \tau_k)$ calculated at a discrete set of pulse durations $t_j$ and free evolution times $\tau_k$. For simplicity, we work with evenly spaced times along both dimensions.

\begin{figure}[t!]
\includegraphics{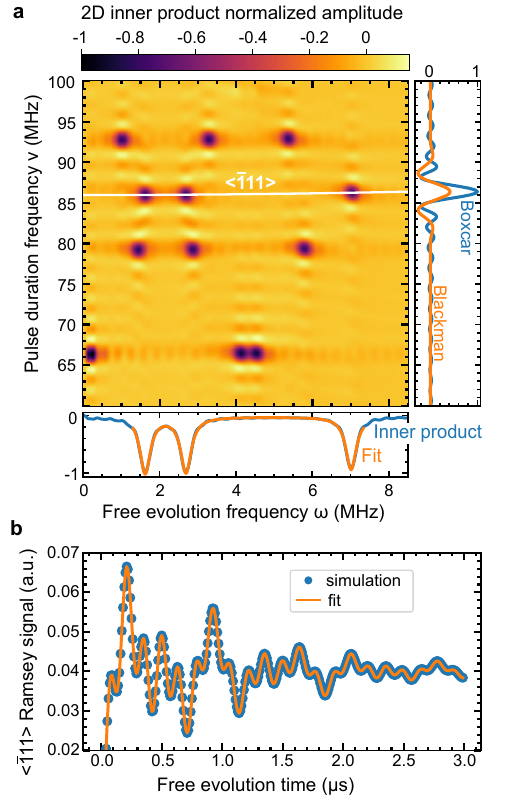}
    \caption{Inner-product-based inversion. All plots in this figure analyze the same simulated $SQ$-cancelled data set $S_\text{VPDR}(t_j, \tau_k)$ performed for $\mathbf{B_\text{DC}} = (-38.4, 25.7, 19.1)~\upmu$T in conventional diamond unit cell coordinates, at the optimum MW angle from Fig.~\ref{fig3}, with $\Omega_\text{max} = 100$ MHz,  $T_2^* = 2\upmu$s, a maximum pulse duration $t_j$ of 397.5 ns and a pulse step size of 2.5 ns. a) Boxcar-windowed two-dimensional inner product $I(\nu, \omega)$ as defined in Eqs.~\ref{eq:I}-\ref{eq:f}. 
    (right inset) Filter functions $F(\nu)\propto\sum_j W(t_j)\cos{\nu t_j}$ for boxcar (blue) and Blackman (orange) windows. (lower inset) $I(\nu, \omega)$ evaluated along the white line in (a), with a Blackman window, fit with three Lorentzians. (b) Blackman-windowed $\tau-$domain Ramsey signal $f(\tau_k) \propto \sum_j S_\text{VPDR}(t_j,\tau_k)W(t_j)\cos{(\Omega_{\langle\bar{1}11\rangle} t_j)}$. The signal is fit with the sum of three decaying sinusoids with commensurate decay times, frequencies constrained to $\Delta\omega = 2|\omega_0 + m_I A|$ for $m_I \in \{-1, 0, 1\}$, and variable phase (due to an inner product at fixed $\Omega$ rather than $\Omega_\text{eff}$).}
    \label{fig4}
\end{figure}

The most straightforward inversion approach 
uses a frequency-domain analysis. Since the $DQ$ term of interest has the form $\cos{\Omega_\text{eff} t} \cos{\Delta \omega  \tau}$ (as can be found by expanding powers in Eq.~\ref{eq:P0}), we consider a 
frequency domain signal $I(\nu,\omega)$ obtained from a double-cosine inner product of $S_\text{VPDR}$ given by
\begin{eqnarray}
    I(\nu, \omega) &=& \frac{\sum_k \left(f(\tau_k, \nu) - \bar{f}(\nu)\right) \cos{\omega \tau_k}}{\sum_k \cos^2\omega \tau_k},\label{eq:I}\\    f(\tau_k, \nu) &=& \frac{\sum_j S_\text{VPDR}(t_j,\tau_k) W(t_j) \cos{\nu t_j}}{\sum_j \cos^2{\nu t_j}},\label{eq:f}
\end{eqnarray}
where we subtract the mean value $\bar{f}(\nu) = \frac{1}{M}\sum_kf(\tau_k, \nu)$ and allow different window functions $W(t_j)$. Figure~\ref{fig4}a shows $I(\nu, \omega)$ for a boxcar window $W(t_j) = 1$, for  a range of $\nu$ restricted to show only the $\nu = \Omega_\text{eff}$ features for each orientation; $\Omega_\text{eff}$ for the $\langle\bar{1}11\rangle$ orientation is highlighted in white.  The spectral peaks show fringing, particularly in the vertical dimension, which occurs because $S_\text{VPDR}(t,\tau)$ does not fully decay over the probed values of $t_j$ and $\tau_k$ (especially in the pulse duration dimension $t_j$). These fringes pose a challenge because they lead to crosstalk between the different orientations' signals: the fringes from one orientation overlap the signal of another. We therefore switch to a Blackman window function~\footnote{$W_\text{Blackman}(t_n) = 0.42 - 0.5 \cos{(2\pi n/N)} + 0.08 \cos{(4\pi n/N)}$ where $N$ is the number of pulse durations $t_n$ in the data set (see scipy.signal.windows.blackman documentation)} that suppresses the fringes at the cost of a slightly wider and shallower central lobe (see Fig.~\ref{fig4}a right inset for the associated filter function). With a Blackman window, we take a cut along the effective Rabi frequency (white line) to obtain the frequency-domain Ramsey signal $I\left(\sqrt{\Omega
^2 + \omega^2}, \omega\right)$ for the $\langle \bar{1}11\rangle$ orientation of interest (lower inset of Fig.~\ref{fig4}a). By fitting with a sum of Lorentzians, 
one might seek to estimate the $m_s = \pm 1$ transition frequency.

However, $I(\Omega_\text{eff}, \omega)$ 
is not exactly the sum of Lorentzians, even for ideal exponentially decaying sinusoidal signals in the time domain. Owing to the finite spacing in free evolution durations $\tau_k$, there is a small deformation near zero frequency that distorts the fit and leads to systematic inaccuracies in the estimated transition frequencies (at the $\sim$10-nT level in the example in Fig.~\ref{fig4}a). We thus consider the time-domain Ramsey signal $f(\tau_k, \nu =\Omega_i)$
where $\Omega_i$ is the Rabi frequency for the orientation of interest. Figure~\ref{fig4}b shows $f(\tau_k, \Omega_{\langle\bar{1}11\rangle})$ for the $\langle\bar{1}11\rangle$ orientation with a Blackman window, fit to the sum of three decaying sinusoids. 
This time-domain analysis provides a significantly better estimate of the transition frequencies~\footnote{The presented analysis using frequencies constrained to $\Delta\omega = 2|\omega_0 + m_I A|$ works well for weak transverse fields relative to $\Delta$. Within earth's field, 
it remains compatible with sub-nT accuracy in our inversions of simulated data. At higher fields, this constraint might need to be relaxed to achieve the same level of accuracy.}. Comparing the extracted $\Delta\omega$ for the highest frequency sinusoid to the exact transition frequency $\Delta\omega_\text{exact}$, the fit in Fig.~\ref{fig4}b has an error of $\Delta\omega - \Delta\omega_\text{exact} \approx 20 $~Hz, limited by residual crosstalk from the other orientations. Converting to an equivalent axial magnetic field error $\Delta B = \frac{\hbar}{2g\mu_B}(\Delta\omega - \Delta\omega_\text{exact})$, we extract $\Delta B \approx 0.35$~nT. 

While the time-domain inversion via fitting $f(\tau_k, \Omega_i)$ offers good accuracy, other approaches may prove superior depending on instrument goals. Acquiring data over a full set of MW pulse durations and free evolution times leads to sensitivity that is roughly $\sim 30\times$ worse than single-$\tau$ DQ Ramsey measurements, depending on acquisition details (see Appendix~\ref{app:sensitivity}). Moreover, acquisition of the 48,000 measurements used for the example in Fig.~\ref{fig4} would require more than a second in typical systems, limiting bandwidth. Acquisition time could be reduced by more than an order of magnitude by sampling only at free evolution times when all of the hyperfine signals rephase, $\tau_s = 2\pi s/(2A)$ for $s \in \mathbb{N}$; such sampling would improve sensitivity but limit dynamic range ($g\mu_B B_z/\hbar < A/2$). Bayesian adaptive schemes~\cite{mcmichael_sequential_2021, bonato_optimized_2016} as well as non-adaptive phase estimation algorithms~\cite{waldherr_high-dynamic-range_2012, nusran_high-dynamic-range_2012} could also be applied in the context of the VPDR protocol. 

In particular, if the goal is to sense small deviations from a known field, which is the context in which sensitivity is usually defined, one would measure only at the optimal free evolution time $\tau_\text{opt}^{(i)}$ for each orientation $i$. In this scenario, the cost to sensitivity of the VPDR protocol is minimal: the sensitivity for measuring a single orientation at $\tau_\text{opt}^{(i)}$ is only $2\sqrt{2}$ worse than standard $DQ$ Ramsey in the hard-pulse limit, and approximately 6$\times$ worse for the conditions considered in Fig.~\ref{fig5}-\ref{fig6} (see Appendix~\ref{app:sensitivity}). Furthermore, even though the evolution time is not optimal for them, some information is gained about other non-targeted orientations in VPDR, such that sequential VPDR measurements at all $\tau_\text{opt}^{(i)}$ hold some advantage over sequential $DQ$ Ramsey measurements. Even in this scenario, the refresh rate is reduced by the number of pulse durations sampled, but it may be possible to devise sparse-sampling approaches in both $t$ and $\tau$ dimensions. Here, we focus on the question of inversion accuracy and leave optimization of sensitivity and refresh rate to future development. 

\subsection{Inversion accuracy}
\begin{figure}[t]
\includegraphics{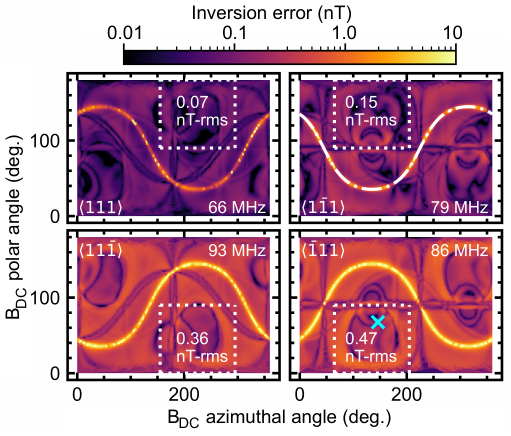}
    \caption{Systematic inversion error for a $B_\text{DC} = 50~\upmu$T field with varying direction relative to diamond crystalline axes. Simulations are run with $\Omega_\text{max}=100$~MHz, $T_2^* = 2\upmu$s, MW direction as shown in Fig.~\ref{fig3}b, pulse durations up to 797.5 ns in 2.5 ns steps, and free evolution times up to 2.98 $\upmu$s in 20 ns steps. The time-domain Ramsey signal $f(\tau, \Omega)$ is calculated for each orientation using a Blackman window and fit to extract the $m_s = \pm 1$ transition frequency of the highest-frequency hyperfine line. The error compared to the exact eigenvalue difference is expressed as equivalent error in the axial field, with plots for each orientation labeled by their crystallographic direction and approximate Rabi frequency $\Omega$. 
    The dashed-dotted white line shows the region of vanishing axial field for the $\langle 1\bar{1}1\rangle$ orientation. Dotted boxes indicate regions in which the indicated rms errors are calculated; the cyan $\times$ marks the field and orientation for which Fig.~\ref{fig4} is calculated.}
    \label{fig5}
\end{figure}

We examine the accuracy of the time-domain inversion 
by simulating and analyzing VPDR signals over a range of magnetic fields. Figure~\ref{fig5} shows the systematic inversion error for a typical Earth magnetic field of 50~$\mu$T, expressed as the equivalent error $\Delta B$ in the axial field along each NV orientation. Sub-nT accuracy is achieved for most DC field angles, with the $\langle 111\rangle$ orientation showing the smallest systematic error because it enjoys the best spectral isolation from the other orientation's Rabi frequencies and harmonics. Away from the large-error contours discussed below, we observe detailed patterns in $\Delta B$ that vary with field magnitude. If we increase the maximum Rabi pulse duration or eliminate non-target orientations from the simulation, these errors are reduced by orders of magnitude, thus demonstrating that they are caused by crosstalk from other orientations. 

Each orientation exhibits zones of poor accuracy when the DC field is perpendicular to its axis; the dashed-dotted line in the $\langle 1\bar{1}1\rangle$ panel of Fig.~\ref{fig5} shows its perpendicular-field contour. These dead zones are associated with vanishing sensitivity of a cosine-like Ramsey signal at zero field, which is ultimately caused by the fixed direction of the applied MW (see Appendix~\ref{app:sensitivity_loss} for further discussion). The loss of sensitivity exacerbates crosstalk effects and leads to poor accuracy. Fortunately, the dead zone angular extent is small; furthermore, the DC field can be reconstructed from three orientations' signals, meaning that true dead zones are only encountered when the DC field is perpendicular to two NV axes simultaneously.

While we only show $50~\upmu$T data, we have also found similar performance in numerical simulations over terrestrial fields $20-70~\upmu$T. The dynamic range is limited by the Nyquist frequency for sampling in $\tau$, which should not exceed $2(\omega_L+A)/2\pi$ for any orientation; for 20~ns sampling, the dynamic range exceeds $400~\upmu$T. 
Even with faster sampling, there will also be an upper limit on range as $\omega_L$ approaches $\Omega_i$: for $\Omega_i/\omega_L < 2$, there is a steady fall-off in sensitivity as the $|\pm\rangle$ states are too far detuned to be effectively driven (see Fig.~\ref{fig2}b). Since $\Omega_\text{max}$ cannot be significantly increased without encountering rotating-wave-approximation limitations, we can estimate an upper limit on 
range of $\omega_L \sim 30 $ MHz or $|B|\sim1$~mT. 

\subsection{Robustness to microwave drift}

\begin{figure}[t]
\includegraphics{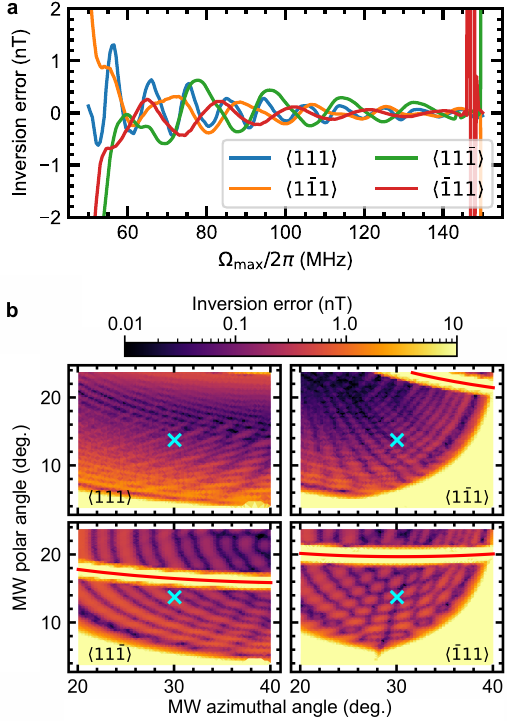}
    \caption{Robustness to drift in MW excitation parameters. (a) Inversion error vs MW amplitude. 
    (b) Inversion error vs MW direction for $\Omega_\text{max}/(2\pi) = 100$~MHz.  Cyan $\times$ marks show the direction used in (a); red lines indicate coincidence of  $3\Omega_{\langle111\rangle}/2$ with $\Omega_i$. Yellow regions exceed 10~nT error.  In  both (a) and (b), simulations are run for $B_\text{DC} = (-38.4, 25.7, 19.1)~\upmu$T (cyan $\times$ in Fig.~\ref{fig5}) with $T_2^* = 2~\upmu$s, pulse durations up to 797.5 ns in 2.5 ns steps, and free evolution times up to $2.98~\upmu$s in steps of 20 ns. Inversions are performed without prior knowledge of precise MW parameters by estimating $\Omega_i$ from $\sum_k f(\tau_k, \nu)$ and then fitting $f(\tau_k, \Omega_i)$ for each orientation.}
    \label{fig6}
\end{figure}

Up until this point, we have performed inversions assuming known MW amplitude and direction. However, a primary strength of the VPDR protocol is that it does not require perfect calibration of the MW field; it is robust to slow drifts in MW amplitude and direction because the Rabi frequency for each orientation can be determined from the signal. In some sense, the labeling of NV orientations is self-calibrating. Qualitatively, by integrating $I(\nu, \omega)$ (see Fig.~\ref{fig4}a) over $\omega$  one obtains a Rabi spectrum whose features permit estimation of $\Omega_i$. In practice, we calculate $\sum_{k} f(\tau_k, \nu)$, which exhibits peaks near $\nu = \Omega_i$. Provided the drifts in MW excitation are sufficiently small that the features retain their ordering, $\Omega_i$ can thereby be estimated for each NV axis. Furthermore, slightly imperfect Rabi-frequency estimation 
does not significantly impact the frequencies in the time-domain Ramsey signal. Thus even fairly large drifts in the MW ``label" do not cause concomitant systematic errors in the magnetic field measurement. 

Figure~\ref{fig6} illustrates the robustness of the VPDR protocol to variation in MW power and direction when estimating Rabi frequencies from simulated data. Inversions in this figure assume a known ordering of Rabi frequencies, corresponding to an approximate MW direction, but do not otherwise employ a priori knowledge of MW parameters. As the MW amplitude increases, the inversion error $\Delta B$ decreases because the different orientations' signals become better separated and crosstalk is reduced (see Fig.~\ref{fig6}a). At some point, the Rabi frequencies become so high that aliasing in the data becomes an issue, leading to the wild variations for $\Omega_\text{max} \gtrsim 145$~MHz; this could be mitigated by reducing the pulse duration step size $\Delta t = t_{j+1} -t_j$ in $S_\text{VPDR}(t_j,\tau_k)$. Regardless, Fig.~\ref{fig6}a illustrates robustness to factor-of-two variations in MW amplitude. Variations in MW direction (see Fig.~\ref{fig6}b) produce acceptable errors over at least $8\degree$ ranges in the polar and azimuthal angles between 
the diamond crystalline axes and the MW field. The large range of tolerable MW angles and magnitudes seen in Fig.~\ref{fig6} is representative of typical results in terrestrial fields, though detailed cross-talk-induced features in the errors vary. In all cases, the angular limits are constrained by the $3\Omega_{\langle 111\rangle}/2$ harmonic, which produces significant crosstalk as it approaches the other orientations' Rabi frequencies (red lines). 

The above analysis concerns quasi-static changes to the MW field, which are an essential consideration for long-term accuracy. MW intensity fluctuations on timescales faster than a measurement could elongate features along the pulse duration frequency axis, potentially increasing crosstalk (see Fig.~\ref{fig4}a). Nevertheless, some noise is tolerable provided that the associated Rabi broadening is small compared to the dephasing rate or the inverse of the maximum pulse duration. It is also worth noting that our analysis implies robustness to drifts in MW detuning due to e.g. temperature fluctuations. Detuning does not impact the double-quantum transition frequency, and the typical changes it induces in effective Rabi frequency are tiny compared to the range that VPDR can tolerate.

Ultimately, these numerical investigations demonstrate the possibility to accurately extract NV spin transition frequencies along the four diamond bond directions, without relying on precision calibration of the MW direction used to label the NV orientations. 

\section{Experimental feasibility study \label{sec:experiment}}
\begin{figure*}[ht]
    \includegraphics{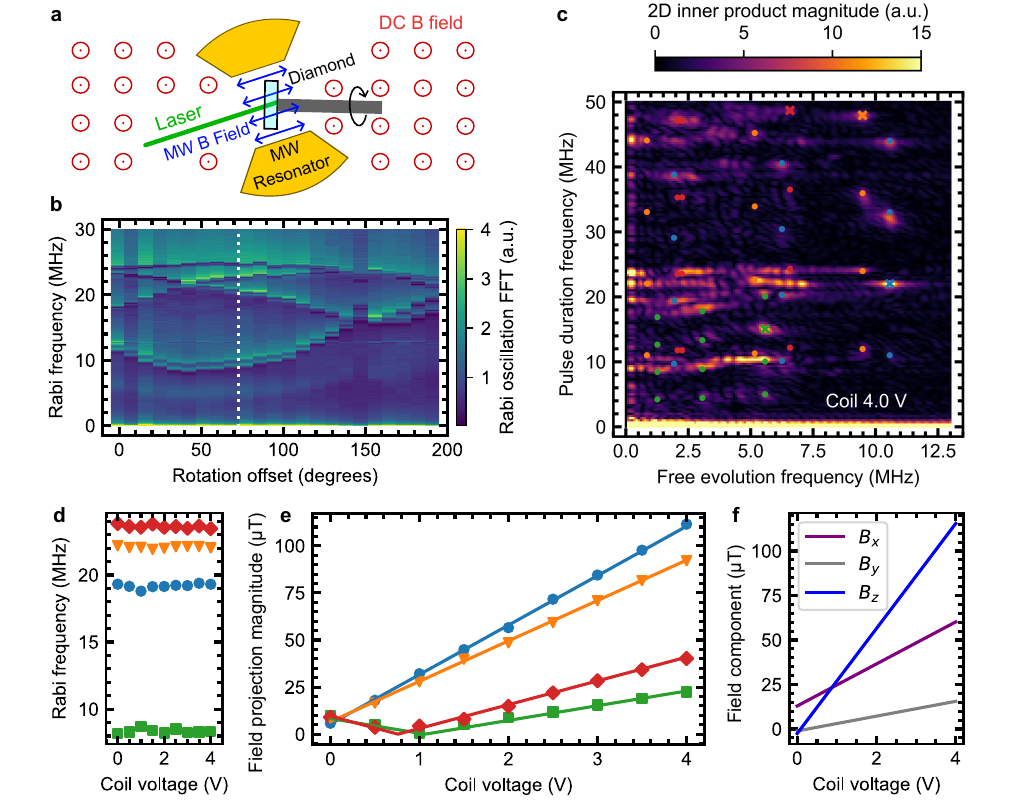}
    \caption{Experimental feasibility study. (a) Top 
    view of device illustrating ability to rotate the diamond relative to the MW magnetic field. 
    (b) Fast Fourier transform of Rabi oscillation data as a function of diamond angle in the absence of applied DC field. The angle chosen for experiments is indicated with the dashed white line. (c) Example Fourier-domain data set at a coil voltage of 4~V. VPDR data are acquired at pulse durations from 50 ns to 1.995~$\upmu$s in 5 ns steps and free evolution times from 50 ns to 2.99~$\upmu$s in 30 ns steps, and averaged over acquisitions at MW frequencies of 2.867, 2.8674, and 2.8678 GHz. 
    Local maxima associated with a $DQ$ signal are marked with a $\times$ and used to infer expected locations of other $DQ$ peaks for the same orientation, indicated by dots of the same color. (d) Inferred bare Rabi frequency found from inner-product maxima ($\times$'s in (c)) as a function of coil voltage. Different symbols are associated to distinct but unknown NV orientations. (e) Inferred projections of the magnetic field on each orientation's axis. 
    Solid lines show field projection magnitudes from a fit to a magnetic field that varies linearly with voltage. (f) Field components, expressed in crystal coordinates, of the linearly varying field fit to the inferred field projections in (e). Note that this field is not unique.  
}
    \label{fig7}
\end{figure*}

To probe the experimental feasibility of realizing the VPDR protocol and verify its qualitative features, we have performed preliminary tests with an improvised device. Figure~\ref{fig7}a illustrates the key capabilities of the system, which was constructed by adapting an existing CW-ODMR magnetometer prototype. The diamond sample is mounted on a rod within a 3D MW resonator~\cite{halde_conception_2021} between two bowtie pillars that determine the direction of the MW field, while permitting relatively high Rabi frequencies in the 10s of MHz. Rotation of the rod enables partial control over the orientation of the diamond crystal axes relative to the driving MW field. NVs in the sample are illuminated with a 520~nm laser (approximately 100 mW), and fluorescence is detected by a photodiode below the diamond. MW pulses are generated with a voltage-controlled oscillator gated by a FPGA-based controller. The entire device can be placed inside a uniaxial solenoid coil within high-permeability shielding to explore variation with external magnetic field. 

As the diamond is rotated relative to the MW resonator, the Rabi frequencies for each orientation vary. Figure~\ref{fig7}b shows the fast Fourier transform (FFT) of the ensemble Rabi oscillation as a function of rotation angle, 
clearly illustrating four distinct Rabi frequencies (split by hyperfine interactions, particularly at lower Rabi frequency). While we can reach angles that adequately separate the bare Rabi frequencies $\Omega_i$ of the four NV orientations, the ideal angle shown in Fig.~\ref{fig3}b is not attainable, and some of the Rabi harmonics overlap. Furthermore, in this prototype we do not have a priori knowledge of the absolute diamond orientation, so we cannot assign Rabi frequencies to specific crystallographic directions. 

At a diamond rotation angle of approximately $73\degree$, we perform the VPDR protocol as a function of voltage sent to the uniaxial coil to verify that anticipated features are observable and move with field as expected. Because the MW delivery system did not have dynamic phase control capabilities, we identify the $DQ$ signals by averaging data over detuning. Unlike $DQ$ signals, $SQ$ signals occur at free evolution frequencies that change with detuning, so the detuning-averaged $SQ$ signals are smeared out horizontally in the Fourier domain, whereas the $DQ$ signals remain point-like.

Figure~\ref{fig7}c shows a sample Fourier-domain data set obtained by averaging over three detunings 
at a solenoid coil voltage of 4 V. The Fourier representation employs a mean-subtracted double inner product given by Eq.~\ref{eq:I}-\ref{eq:f} with complex exponentials replacing the cosines~\cite{koppens_universal_2007}, using a cosine window function.  
Local peaks in the Fourier-domain VPDR data at large free-evolution frequencies can be attributed to $DQ$ signals from each of the four NV orientations. $\times$'s mark local extrema associated with an $m_I = +1$, $DQ$ feature from each orientation. We find a local maximum in the inner product magnitude to obtain 
the extrema locations $(\omega_\text{max,i}, \nu_\text{max,i})$ along with their inferred harmonic $n$ (where $\nu_\text{max,i} = n \Omega_\text{eff,i}$), and thereby 
extract the bare Rabi frequency $\Omega_i = \sqrt{\nu_\text{max,i}^2/n^2 - \omega_\text{max,i}^2}$, approximate field projection $|B_{z,i}| = \hbar|\omega_\text{max,i} - 2A|/(2g\mu_B)$, and expected locations of other $DQ$ features from the same orientation (marked with dots of the same color). While the agreement is not perfect, we do observe features at or near many of the expected locations, particularly for the orientations marked in green, blue and orange. The poor visibility of the fourth orientation (red) could be explained by optical polarization selection rules. The numerous  peaks at moderate free evolution frequency are attributed to $SQ$ signals, which are horizontally tripled owing to the three detunings over which we average the data.

We perform the analysis illustrated in Fig.~\ref{fig7}c for a range of solenoid coil voltages, extracting the Rabi frequency (Fig.~\ref{fig7}d) and axial field projection (Fig.~\ref{fig7}e) for each orientation. The Rabi frequencies $\Omega_i$ remain relatively constant with voltage, and occur at values compatible with the Rabi FFT data considering several days elapsed time between the Rabi and VPDR datasets (see Fig.~\ref{fig7}b dotted white line). In contrast, the field projections evolve significantly with coil voltage. To determine whether our observations are consistent with a uniaxial field that increases linearly with coil voltage, we fit our data to field projections from a linear model (see Fig.~\ref{fig7}f and Appendix~\ref{app:experimental}). With no absolute knowledge of the diamond orientation in our resonator, we could not unambiguously match the ordering of Rabi frequencies to an ordering of known NV orientations, although such identification could be achieved by calibration in triaxial coils~\cite{igarashi_tracking_2020, wang_simultaneous_2025}. We also do not know the sign of the field projection on each NV axis (which could be determined by pulsing a small magnetic field with approximately known direction). 
As a consequence, the inversion in Fig.~\ref{fig7}e-f is not unique (different permutations and signs of the field 
projections can yield the same result), but the good correspondence between the fit and data indicates that our measurements follow expected linear behavior and that the projection on the fourth orientation is compatible with the other three. The offset in field at zero coil voltage can be explained by magnetic components in the rotation stage holding the diamond. 

While the qualitative features of the data behave as expected, we leave quantitative tests of accuracy to future experiments. In Fig.~\ref{fig7}e, we observe rms deviations of 0.7~$\upmu$T between the data and fit, but these likely arise from non-idealities of our apparatus. The coil was not calibrated, and the voltage source used has an accuracy of 0.05\%+10 mV, corresponding to 0.3-0.4~$\upmu$T over the field range inferred from the fit. Furthermore,  the magnetometer used for these tests presented additional sources of systematic error. The diamond support rod blocked the access path for detecting transmitted laser light, which normally would be used as a reference to account for fluctuations in laser power; there was also observable electronic crosstalk, mechanical vibration, and thermal drift. These issues added significant 
systematic uncertainty, but they are straightforward to remedy. Similarly, adding a phase-control circuit would suppress the $SQ$ features that complicate the presented analysis.
Nevertheless, even with poor data quality, it was possible to interpret VPDR features and extract magnetic field estimates that are consistent with the linearly increasing applied field, providing a proof-of-principle demonstration of the VPDR protocol.  

One notable shortcoming of our device was its microwave resonator, which was not purpose-designed and could only reach $\sim$30 MHz Rabi frequency. Future devices could be optimized to increase both MW strength and homogeneity. 
While planar resonators have reached 165 MHz Rabi frequencies~\cite{mariani_system_2020}  and coil-based MW antenna designs can exhibit 60 MHz Rabi frequency over a mm$^3$ homogeneous volume~\cite{park_design_2021}, 3D re-entrant cavities with bow-tie posts~\cite{halde_conception_2021} offer excellent homogeneity and spatial field confinement, and we anticipate that careful optimization of post geometry~\cite{goryachev_high-cooperativity_2014} will permit reaching 100 MHz. Indeed, the fundamental limit on Rabi frequency set by the resonator $Q$, frequency $f$, effective volume $V$, and input power $P_\text{in}$ permits Rabi frequencies $\sim$ 330 MHz at $Q = 100, V = ($1~mm$)^3$, $P_\text{in}=10~$W~\footnote{$Q$ = 100 induces a rise time $\sim$5~ns, but by discarding the first few MW pulse durations (or not acquiring it), our simulations show its impact can be mitigated.}. Additionally, our simulations show sub-nT rms errors for $\Omega_\text{max}$ down to 60 MHz. Taken together, these considerations show the feasibility of future quantitative experimental tests of VPDR accuracy. 

\section{Conclusion}

The VPDR protocol offers a new paradigm for bias-field-free vector NV magnetometry, using Rabi labeling to 
isolate the signal from different orientations in an NV ensemble with minimal crosstalk. Our analytic modeling provides a resource for understanding and interpreting VPDR data, as well as identifying optimal regimes of operation. Numerical simulations of the protocol demonstrate its compatibility with high-accuracy field estimation, even when the MW field used to label NV orientations is only approximately known. Practically, the scheme requires only a single MW frequency, and is not sensitive to slow changes in detuning. 
The primary design requirement for experimental implementation is a relatively strong and homogeneous MW drive field, 
and VPDR data could be measured and interpreted even with an improvised device. The protocol avoids the use of additional hardware, and permits bias-field-free vector field measurements in a compact geometry.

There remain some challenges to working with the VPDR protocol instead of a bias field. A mechanism to determine the sign of the magnetic field projection on each NV axis is still required, e.g. by pulsing a coil to add a known magnetic field shift and observe how each orientation's signal changes. Note that the direction and magnitude of such a pulsed field need only be known approximately to obtain the sign information. In addition, sensitivity dead zones will occur where the external field is perpendicular to two NV axes, requiring multiple sensors or piezo actuation, similar to designs for atomic magnetometers~\cite{leger_-flight_2015, geometrics_how_nodate}. Finally, the simple data acquisition and inversion proposed here precludes high-bandwidth sensing, though it is likely that faster approaches exist. 

Indeed, there are opportunities for refinement of the VPDR protocol as well as its numerical modeling and experimental implementation. MW resonator design targeting power and homogeneity would greatly simplify data interpretation and inversion. More sophisticated theory including internal electric fields~\cite{mittiga_imaging_2018, yu_optically_2024} and non-secular hyperfine terms~\cite{felton_hyperfine_2009} may be required for high-accuracy interpretation of experimental data in low fields. Improved bandwidth and sensitivity can likely be achieved with different sampling approaches and inversion methods, and it may be advantageous to incorporate existing sensitivity-enhancement techniques such as P1 driving~\cite{bauch_ultralong_2018}. It may also be possible to adapt our approach to other defect magnetometer technologies such as the $V_\text{Si}$ in SiC~\cite{niethammer_vector_2016}. 
With a recent surge of interest in low-field~\cite{vetter_zero-_2022, wang_zero-field_2022, zheng_zero-field_2019} and high-accuracy~\cite{lonard_limits_2025, tsukamoto_accurate_2022} NV magnetometry, VPDR provides a timely opportunity to eliminate bias fields and improve the accuracy of NV vector-field sensors. 


\begin{acknowledgments}
We thank Fr\'ed\'erik Coulombe and William P\'epin for assistance with the mechanical and electrical modifications of the magnetometer prototype to enable VPDR functionality. We acknowledge contributions of Zackary Flansberry for early efforts on Rabi labeling of NV orientations, and the work of Olivier Bernard,  Hubert Dub\'e, Benjamin Dupuis, Guillaume Duclos-Cianci, Nicolas Fleury, Fran\c{c}ois Ouellet, Amin Resaei, and Gabriel St-Hilaire 
in developing the original magnetometer prototype. 
This work was supported by the Minist\`ere de l'\'Economie et de l'Innovation (MEIE) grant PI70865 and National Research Council (NRC) grant QSP-051. 
\end{acknowledgments}

\vspace{.2in}
\centerline{\bf CONFLICT OF INTEREST}
\vspace{.2in}
L.C. reports paid employment (half-time) with SBQuantum from February 15, 2024-February 14, 2025. L.C., V.H., K.J., A.L., D. R.-G., R.R., and A.S. are named inventors of the provisional patent application on the VPDR protocol examined in this article; the provisional patent is owned by SBQuantum. D. R.-G. and V.H. hold equity in SBQuantum.

\appendix

\section{\label{app:analytic} Analytic model at finite $\Omega/\omega_L$ and with dephasing}

The probability for an $m_s = 0$-prepared spin to end up in $m_s=0$ after application of the VPDR pulse sequence is given by $P_0(t,\tau)$. For a coherently-evolving single NV center in an axial magnetic field, we can analytically calculate this probability by solving the Schr\"{o}dinger equation with a piecewise-constant Hamiltonian given by $H_1$ from time $x = 0$ to $x = t$, $H_0$ from $x = t$ to $x = t + \tau$, and $H_{1\phi}$ from $x = t+\tau$ to $x = 2t + \tau$. $H_1, H_0$ and $H_{1\phi}$ are derived from $H$ given in Eq.~\ref{eq:H} as follows: In the $\Omega \rightarrow \infty$ limit, we set $H_1 = H(\phi = 0, \delta = 0,\omega_L = 0), H_0 = H(\Omega = 0),$ and $H_{1\phi} = H(\delta = 0, \omega_L = 0)$, for which simple analytic solutions can be found. For finite $\Omega/\omega_L$ calculations, we set $H_1 = H(\phi = 0, \delta = 0), H_0 = H(\Omega = 0, \delta = 0),$ and $H_{1\phi} = H(\delta = 0)$, allowing for finite $\omega_L$ during all time segments; we restrict to zero detuning because it keeps the expressions relatively simple and reflects conditions under which the protocol is likely to be run. 

At finite $\Omega/\omega_L$ and zero detuning $\delta = 0$, the VPDR signal for a single NV without hyperfine structure comprises oscillations at $\nu_n = n\sqrt{\Omega^2 + 4 \omega_L^2} = n\Omega_\text{eff}$ and $\omega_m = m \omega_L$ in $t$ and $\tau$ respectively, where $n = \{-2, -\frac{3}{2},-1, -\frac{1}{2}, 0, \frac{1}{2}, 1, \frac{3}{2}, 2\}$ and $m = \{-2, -1, 0, 1, 2\}$. In the absence of dephasing, the $m_s =0$ population after the VPDR pulse sequence can be found analytically and expressed as
\begin{equation}
    P_0(t,\tau) = \sum_n\sum_m a_{n,m}e^{i\nu_n t}e^{i\omega_m \tau}.
    \label{eq:P0extended}
\end{equation} The coefficients $a_{n,m}$ depend only on the ratio $\alpha = \Omega/\omega_L$ and the relative phase $\phi$ of the second MW pulse, and $a_{n,m}$ are given in Table~\ref{tab:coeff} for $n \geq 0$. Since $P_0(t,\tau)$ is real-valued, the remaining coefficients can be found from the relationship $a_{-n, -m} = a_{n, m}$.

\begin{sidewaystable}
\centering
\begin{tabular}{l|ccccc}
     n & $m=-2$ &$m= -1$ & $m=0$ & $m=1$ & $m=2$ \\
     \hline
\noalign{\vskip 1mm} 
2  & $\frac{\alpha ^8 \left(\tilde{\alpha }-8\right)+32 \alpha ^6 \left(\tilde{\alpha }-3\right)+128 \alpha ^4 \left(\tilde{\alpha
   }-2\right)}{64 \tilde{\alpha }^9} $& $\frac{\alpha ^6 \left(\alpha ^2 \left(\tilde{\alpha }-4\right)+8 \left(\tilde{\alpha }-2\right)\right) \cos (\phi )}{16 \tilde{\alpha
   }^9}$ & $\frac{3 \alpha ^8}{32 \tilde{\alpha }^8}$& $\frac{\alpha ^6 \left(\alpha ^2 \left(\tilde{\alpha }+4\right)+8 \left(\tilde{\alpha }+2\right)\right) \cos (\phi )}{16 \tilde{\alpha
   }^9}$ & $\frac{\alpha ^8 \left(\tilde{\alpha }+8\right)+32 \alpha ^6 \left(\tilde{\alpha }+3\right)+128 \alpha ^4 \left(\tilde{\alpha
   }+2\right)}{64 \tilde{\alpha }^9}$ \\
$\frac{3}{2}$  & $\frac{\alpha ^6 \left(20-6 \tilde{\alpha }\right)-32 \alpha ^4 \left(\tilde{\alpha }-2\right)+\alpha ^8}{4 \tilde{\alpha }^9}$ & $\frac{\alpha ^4 \left(16 \left(\tilde{\alpha }-2\right)+\alpha ^4-4 \alpha ^2\right) \cos (\phi )}{2 \tilde{\alpha }^9}$ &$\frac{3 \alpha ^6}{\tilde{\alpha }^8}$ & $\frac{\alpha ^4 \left(16 \left(\tilde{\alpha }+2\right)-\alpha ^4+4 \alpha ^2\right) \cos (\phi )}{2 \tilde{\alpha }^9}$ & $-\frac{\alpha ^4 \left(\alpha ^2 \left(6 \tilde{\alpha }+20\right)+32 \left(\tilde{\alpha }+2\right)+\alpha ^4\right)}{4 \tilde{\alpha
   }^9}$ \\
1  & $-\frac{\alpha ^4 \left(\alpha ^2-24\right) \left(\alpha ^2 \left(\tilde{\alpha }-4\right)+8 \left(\tilde{\alpha }-2\right)\right)}{16
   \tilde{\alpha }^9}$ & $-\frac{\alpha ^2 \left(-4 \alpha ^4 \left(\tilde{\alpha }+3\right)+8 \alpha ^2 \left(3 \tilde{\alpha }-4\right)-64 \left(\tilde{\alpha
   }-2\right)+\alpha ^6\right) \cos (\phi )}{2 \tilde{\alpha }^9}$ &$\frac{\alpha ^4 \left(\alpha ^4-16 \alpha ^2+256\right)}{8 \tilde{\alpha }^8}$ & $\frac{\alpha ^2 \left(4 \alpha ^4 \left(\tilde{\alpha }-3\right)-8 \alpha ^2 \left(3 \tilde{\alpha }+4\right)+64 \left(\tilde{\alpha
   }+2\right)+\alpha ^6\right) \cos (\phi )}{2 \tilde{\alpha }^9}$ & $-\frac{\alpha ^4 \left(\alpha ^2-24\right) \left(\alpha ^2 \left(\tilde{\alpha }+4\right)+8 \left(\tilde{\alpha }+2\right)\right)}{16
   \tilde{\alpha }^9}$ \\
$\frac{1}{2}$  & $-\frac{\alpha ^4 \left(3 \alpha ^2-16\right) \left(-2 \tilde{\alpha }+\alpha ^2+4\right)}{4 \tilde{\alpha }^9}$ & $\frac{\alpha ^2 \left(-4 \alpha ^4 \left(2 \tilde{\alpha }+1\right)+16 \alpha ^2 \left(3 \tilde{\alpha }+2\right)-128 \left(\tilde{\alpha
   }-2\right)+\alpha ^6\right) \cos (\phi )}{2 \tilde{\alpha }^9}$ &$\frac{\alpha ^2 \left(5 \alpha ^4-32 \alpha ^2+128\right)}{\tilde{\alpha }^8}$ & $-\frac{\alpha ^2 \left(\alpha ^4 \left(8 \tilde{\alpha }-4\right)+\alpha ^2 \left(32-48 \tilde{\alpha }\right)+128 \left(\tilde{\alpha
   }+2\right)+\alpha ^6\right) \cos (\phi )}{2 \tilde{\alpha }^9}$ & $\frac{\alpha ^4 \left(3 \alpha ^2-16\right) \left(2 \tilde{\alpha }+\alpha ^2+4\right)}{4 \tilde{\alpha }^9}$ \\
0  & $\frac{\alpha ^4 \left(3 \alpha ^4-96 \alpha ^2+128\right)}{32 \tilde{\alpha }^8}$ & $-\frac{\alpha ^2 \left(\alpha ^6-24 \alpha ^4+320 \alpha ^2-512\right) \cos (\phi )}{8 \tilde{\alpha }^8}$ &$\frac{9 \alpha ^8+64 \alpha ^6+1536 \alpha ^4+4096}{16 \tilde{\alpha }^8}$ & $-\frac{\alpha ^2 \left(\alpha ^6-24 \alpha ^4+320 \alpha ^2-512\right) \cos (\phi )}{8 \tilde{\alpha }^8}$ & $\frac{\alpha ^4 \left(3 \alpha ^4-96 \alpha ^2+128\right)}{32 \tilde{\alpha }^8}$ \\
\end{tabular}
\caption{The coefficients $a_{n,m}$ for  $P_0(t,\tau) = \sum_n\sum_m a_{n,m}e^{i\nu_n t}e^{i\omega_m \tau}$ for a single NV without hyperfine structure at zero detuning and with no dephasing. Expressions are given in terms of $\alpha = \Omega/\omega_L$ and $\tilde{\alpha} = \sqrt{\alpha^2 + 4}$ for concision.}
\label{tab:coeff}
\end{sidewaystable}

To include dephasing in the model in an analytically tractable manner, we neglect spin dephasing during the pulses because we expect $\max(t) \ll \max(\tau)$ in experiments employing high Rabi frequency $\Omega\gg 1/T_2^*$. We also assume that the spin dephasing is quasi-static and magnetic in nature, 
and we model it using a random internal axial magnetic field $B_z^0$ parameterized by a Lorentzian probability distribution of the associated Larmor frequencies 
$g(\omega_L^0) = \frac{T_2^*}{\pi  \left((T_2^*)^2 (\omega_L^0)^2+1\right)}$
~\cite{dobrovitski_decoherence_2008}.  In this quasi-static model, each spin evolves according to Eq.~\ref{eq:P0extended} with a Larmor precession frequency $\omega_L + \omega_L^0$ with $\omega_L^0$ randomly chosen from the distribution $g(\omega_L^0)$, such that the total average probability of observing the spins in $m_s = 0$ is the convolution of $P_0(t, \tau)$ (as a function of $\omega_L$) and $g$. 
The average signal is thereby given by
\begin{eqnarray}
\overbar{P_0}(t,\tau) &=& \sum_n\sum_m a_{n,m}e^{i n\Omega_\text{eff} t}\int_{-\infty}^\infty g(x - \omega_L) e^{i m x \tau} dx \nonumber\\
 &=& \sum_n\sum_m a_{n,m}e^{-|m|\tau/T_2^*}e^{i n \Omega_\text{eff} t}e^{i m \omega_L \tau},
\end{eqnarray}
where $n$ and $m$ are summed over the ranges given for Eq.~\ref{eq:P0extended}. Here we have neglected the impact of $\omega_L^0$ on $\Omega_\text{eff}$, which reflects our neglect of dephasing during the pulses and is an approximation appropriate for $\Omega\gg \omega_L, 1/T_2^*$. 

Note that even in the finite-$\Omega/\omega_L$ regime with dephasing, the coefficients of the $SQ$ terms (corresponding to $m = \pm 1$) are proportional to $\cos{\phi}$, indicating that the $SQ$ terms can be eliminated by addition of two signals acquired with (1) $\phi = 0$ and (2) $\phi = 180\degree$ relative phases for the MW pulses. 

\section{\label{app:sensitivity} Sensitivity analysis}
\subsection{\label{app:senscomparison} Comparison to standard double-quantum Ramsey measurements}
\subsubsection{Hard-pulse limit $\Omega/\omega_L \rightarrow \infty$}

Here, we analyze the sensitivity cost to the VPDR protocol relative to a more typical double-quantum Ramsey experiment performed at fixed pulse duration $t = \pi/\Omega$ in near-zero field. For simplicity, we imagine a situation where there is only a single NV orientation since low-field ``standard" $DQ$ Ramsey would only work in this scenario. We assume that we are operating in the $\Omega/\omega_L \rightarrow \infty$ limit such that simple analytic expressions apply, and we assume the readout is characterized by Gaussian noise independent of readout value (as appropriate for low-contrast fluorescence readout). Since sensitivity is usually defined in the context of a known initial field (from which one senses a small change), we consider measurements at a single free evolution time $\tau$ that is the same for both DQ Ramsey and VPDR schemes. 

For the VPDR measurement, we consider an experiment comprising $N\gg 1$ measurements of $P_0(t_j, \tau)$ at pulse durations $t_0,\dots t_{N-1}$, producing a fluorescence signal $S = a + b P_0(t_j, \tau)$ where $a$ and $b$ depend on the details of the measurement scheme.  The signal is analyzed via an inner product with $\cos{\Omega t_j}$, possibly with a windowing function $W(t_j)$ applied. This procedure defines the analyzed signal
\begin{equation}
f_{\text{VPDR}}(\tau) = \frac{\sum_{j=0}^{N-1} W(t_j) \cos{\Omega t_j}(a + b P_0(t_j, \tau))}{\sum_{j=0}^{N-1} \cos^2{\Omega t_j}}
\label{eq:fvpdr}
\end{equation}
whose uncertainty can be found via standard Gaussian uncertainty propagation as
\begin{equation}
\Delta f_\text{VPDR} = \Delta S\frac{\sqrt{\sum_{j=0}^{N-1}{W^2(t_j) \cos^2{\Omega t_j}}}}{\sum_{j=0}^{N-1} \cos^2\Omega t_j},
\end{equation}
where $\Delta S$ is the uncertainty in each fluorescence measurement (assumed constant as appropriate for low contrast). 
In the limit that the $t_j$ are evenly spaced over a range much larger than $2\pi/\Omega$ with a spacing much smaller than $2\pi/\Omega$, these sums can be approximated by integrals, such that $\sum_{j=0}^{N-1} \cos^2\Omega t_j \rightarrow \frac{N}{2}$. If we can also assume that $W(t)$ varies slowly compared to the oscillations at frequency $\Omega$, then in the numerator we can divide the sum up into $M$ chunks where $W(t)$ is approximately constant during each chunk, and the sinusoidal signals can still be approximated by their average values integrated over many oscillations. In particular, if $t_j = j \Delta t$ and $M$ evenly divides $N$, this chunking would look like
\begin{eqnarray}
    &\sum_{j=0}^{N-1} W(t_j) \cos{\Omega t_j}(a + b P_0(t_j, \tau))  \approx&  \nonumber\\&\sum_{j=0}^{M-1} W(t_{j\frac{N}{M}})\underbrace{\sum_{i=0}^{N/M-1}\cos{\Omega t_{j\frac{N}{M} +i}}(a + b P_0(t_{j\frac{N}{M} +i}, \tau))}_{\left(\frac{N}{M}\right)\frac{b}{4}\sin^2{\omega_L\tau}},\nonumber&
\end{eqnarray}
where we have used 
\begin{equation}
\lim_{T\rightarrow\infty}\frac{1}{2T}\int_{-T}^T dt \cos{\Omega t} P_0(t, \tau) = \frac{1}{4}\sin^2{\omega_L \tau}
\end{equation}
(in the $\Omega / \omega_L \rightarrow \infty$ limit) to approximate the inner sum. Since the inner sum no longer depends on $j$, we can approximate the outer sum by $M \overbar{W}$, where  $\overbar{W}$ is the mean value of the windowing function. Ultimately, including the factor of $2$ from the denominator of Eq.~\ref{eq:fvpdr}, we obtain
\begin{equation}
    f_\text{VPDR}(\tau)\approx \frac{b}{2} \overbar{W} \sin^2{\omega_L \tau},
\end{equation}
and, by similar reasoning,
\begin{equation}
    \Delta f_{VPDR} \approx \Delta S\sqrt{\frac{2}{N}}\sqrt{\overbar{W^2}}.
\end{equation}

We compare this analyzed VPDR signal to the signal obtained from averaging $N $ repetitions of a standard double-quantum Ramsey experiment,
\begin{eqnarray}
f_R &=& a + b \cos^2{\omega_L \tau}     \\
\Delta f_R &=& \Delta S\frac{1}{\sqrt{N}}
\end{eqnarray}

The sensitivity of each experiment is given by 
\begin{equation}
\eta = \Delta B \sqrt{T} = \frac{\hbar}{g\mu_B}\Delta \omega_L \sqrt{T} = \frac{\hbar}{g\mu_B} \frac{\Delta f}{|df/d\omega_L|} \sqrt{T},
\end{equation}
where $T$ is the total time required for an experiment that results in magnetic field uncertainty $\Delta B$.
For $\Omega \gg \omega_L$, the time required for the Rabi pulses is negligible compared to the free evolution and measurement time, so we can approximate that $T$ is the same for both protocols. Thus we can compare the sensitivity of the VPDR and double-quantum Ramsey experiments by finding the ratio of their signal to noise:
\begin{eqnarray}
    \frac{\eta_\text{VPDR}}{\eta_R} &=& \frac{\Delta f_\text{VPDR}}{|df_\text{VPDR}/d\omega_L|}\frac{|df_R/d\omega_L|}{\Delta f_R}\\
    &\approx& \frac{\Delta S \sqrt{\overbar{W^2}}\sqrt{2/N}}{b \overbar{W} \tau |\sin{2\omega_L \tau}|/2} \frac{b~\tau |\sin{2\omega_L\tau}|}{\Delta S /\sqrt{N}}\\
    &\approx& 2 \sqrt{2}\frac{\sqrt{\overbar{W^2}}}{\overbar{W}}.\label{eqn:sensitivityratio}
\end{eqnarray}
If no windowing function is needed, we attain the best VPDR sensitivity at $2\sqrt{2}$ times the standard double-quantum Ramsey sensitivity. 

 Eq.~\ref{eqn:sensitivityratio} is derived in the $\Omega/\omega_L \rightarrow \infty$ limit, and as the weight of the relevant VPDR Fourier peaks varies with $\Omega/\omega_L$, so too will the sensitivity. If we consider finite $\Omega$ at $\delta = 0$, with $SQ$ subtraction, analyzed by an inner product with $\cos{\Omega_\text{eff}}$, the only modification to the above analysis is a reduced weight of the inner-product-extracted $DQ$ Ramsey signal, yielding
\begin{equation}
    \frac{\eta_\text{VPDR}}{\eta_R} \approx 2 \sqrt{2}\frac{\sqrt{\overbar{W^2}}}{\overbar{W}}\frac{\left(\alpha ^2+4\right)^4}{|\alpha ^4 \left(\alpha ^4-16 \alpha ^2-192\right)|},
    \label{eqn:sensitivityratio2}
\end{equation}
where $\alpha = \Omega/\omega_L.$ Unsurprisingly, Eq.~\ref{eqn:sensitivityratio2} is well approximated by Eq.~\ref{eqn:sensitivityratio} for $\Omega/\omega_L \gtrsim 10$.


\subsubsection{Numerical simulations of sensitivity}
Eq.~\ref{eqn:sensitivityratio} holds in the presence of dephasing during free evolution, but the sensitivity is also affected by dephasing during the MW pulses. To elucidate this effect, we performed Monte Carlo estimates of the $\eta_\text{VPDR}/\eta_R$ ratio as a function of the maximum MW pulse duration employed in the VPDR protocol, assuming that both VPDR and DQ Ramsey experiments were performed at the same optimal free evolution time $\tau_\text{opt}$. Specifically, we simulated the SQ-cancelled signal $S_\text{VPDR}(t_j,\tau_\text{opt})$ for a single NV orientation (such that standard low-field DQ Ramsey will also work) both with and without hyperfine structure, and extracted $f_\text{VPDR}(\tau_\text{opt})$ (via Eq.~\ref{eq:f}) and $f_R(\tau_\text{opt}) = S_\text{VPDR}(t_\pi, \tau_\text{opt})$. We calculated the slopes $df/d\omega_L$ for each by performing the simulation at slightly different magnetic fields (differing in magnitude by 1 nT). Then, we added randomly generated Gaussian noise with standard deviation $\sigma = 10^{-4}$ to $S_\text{VPDR}(t,\tau_\text{opt})$ ($\sigma$ corresponds to few-nT field excursions, well within the linear regime), from which we obtain a noisy value of $f_\text{VPDR}(\tau_\text{opt})$ via Eq.~\ref{eq:f}. The uncertainty $\Delta f_\text{VPDR}$ is estimated by the standard deviation of our results. The corresponding uncertainty in $f_R(\tau_\text{opt})$ is $\sigma/\sqrt{N}$, where $N$ is the number of MW pulse durations employed to calculate $f_\text{VPDR}(\tau_\text{opt})$, effectively averaging over $N$ noisy measurements of $S_\text{VPDR}(t_\pi,\tau_\text{opt})$. 
The uncertainty in $f_\text{VPDR}(\tau_\text{opt})$ and $f_R(\tau_\text{opt})$ is converted to uncertainty in magnetic field by dividing by the respective slope $df/d\omega_L$, yielding $\Delta B^\text{opt}_p$ for each protocol $p \in \{\text{VPDR}, \text{R}\}$. The resulting ratio of magnetic field uncertainties $\Delta B^\text{opt}_\text{VPDR}/\Delta B^\text{opt}_\text{R}$ is a good approximation to the axial field sensitivity ratio $\eta_\text{VPDR}(\tau_\text{opt})/\eta_R(\tau_\text{opt})$ in the limit that the spin initialization, readout and free evolution take significantly more time than the MW pulses (which is well satisfied in typical ensemble experiments). 

As an example (see Fig.~8a), we show Monte-Carlo estimates of $\eta_\text{VPDR}(\tau_\text{opt})/\eta_R(\tau_\text{opt})$ for the $\langle 1\bar{1}1\rangle $ orientation with $T_2^* = 2~\upmu$s, detuning $\delta = 0$, the same DC magnetic field as Fig.~\ref{fig4} and \ref{fig6}, and the same MW field as Fig.~\ref{fig4} and \ref{fig5}; the $\langle1\bar{1}1\rangle$ orientation's 79.2 MHz Rabi frequency yields a $\pi$ pulse of 6.32 ns, and we used $\tau_\text{opt}=$ 862 ns for $m_I = 0$ only and $\tau_\text{opt}=$ 1.157 $\upmu$s when all three hyperfine lines are included (both $\tau_\text{opt}$ values are found from analytic expressions for the ideal Ramsey signal in the hard-pulse limit; numeric optimization of simulated slopes yielded similar sensitivity ratios). For VPDR, the pulse durations $t_j$ were evenly spaced by $2.5$ ns, and included all $t_j$ strictly less than the maximum pulse duration on the x-axis of Fig.~8a. This analysis was repeated for both boxcar and Blackman window functions.

\begin{figure}[htb]
    \includegraphics{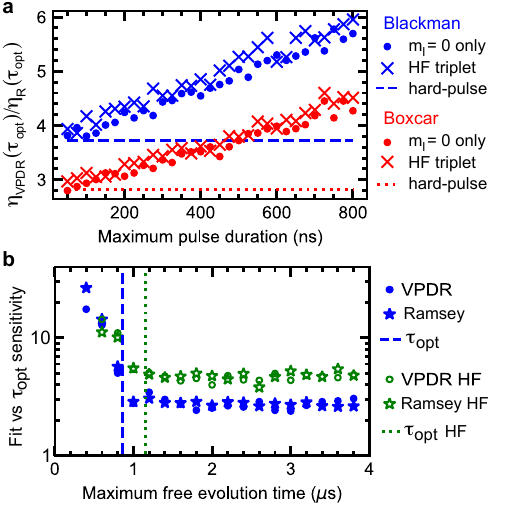}\label{fig8}
    \caption{Sensitivity cost associated with multiple MW pulse durations and multiple free evolution times. Simulations are performed for the same magnetic field as in Fig.~\ref{fig6}, and a MW drive as in Fig.~\ref{fig5}, considering solely the $\langle1\bar{1}1\rangle$ orientation. See text for additional details. (a) Sensitivity cost of many MW durations. Monte-Carlo estimates of $\eta_\text{VPDR}(\tau)/\eta_{R}(\tau)$ for $\tau = \tau_{\text{opt}}$ are shown as a function of the maximum pulse duration used in VPDR, both with (blue) and without (red) a Blackman window on the pulse duration dimension, and both with ($\times$) and without (circles) hyperfine (HF) structure. Dashed and dotted lines show the hard-pulse limit with and without a Blackman window, respectively.  (b) Sensitivity cost of many free evolution times. For each protocol, Monte Carlo simulations estimate the ratio of the sensitivity from fitting signals $f_p(\tau_1, \tau_2,\dots)$ obtained at free evolution times $\tau_k < \tau_\text{max}$ to the sensitivity at the optimal free evolution time $\eta_p(\tau_\text{opt})$, where $p\in\{\text{VPDR}, \text{R}\}$. Evolution both with (green) and without (blue) hyperfine structure  is considered. 
    Dashed (dotted) lines show $\tau_\text{opt}$ in the absence (presence) of hyperfine structure.}
\end{figure}

Figure 8a quantifies the sensitivity cost of sampling at many MW pulse durations in the presence of dephasing. While the sensitivity ratio approaches the hard-pulse limit $2 \sqrt{2}\sqrt{\overbar{W^2}}/\overbar{W}$ when the pulse durations are much shorter than $T_2^*$, for the long pulse durations up to 800 ns included in the analysis of Fig.~\ref{fig5}-\ref{fig6}, we observe a sensitivity cost that is roughly 50\% higher.  Similar results are observed with and without hyperfine structure; we also see similar results for different NV orientations, magnetic fields, and MW pulse step sizes. 
This sensitivity ratio 
is an apples-to-apples comparison of VPDR and DQ Ramsey, both under the same assumption of approximately known external field such that single-$\tau$ measurements are meaningful. 

For the full 2D spectral analysis performed for Fig.~\ref{fig4}-\ref{fig7}, the VPDR sensitivity is worsened by an additional factor arising from sampling at many non-optimal free evolution times $\tau_k$. Fig.~8b shows an example of this factor as a function of the maximum free evolution time, assuming evenly-spaced $\tau_k$ starting from 0 with $\tau_k < \tau_\text{max}$. We again consider the $\langle1\bar{1}1\rangle$ orientation with the same parameters as in Fig.~8a and a Blackman window function. We add Gaussian noise to $S_\text{VPDR}(t,\{\tau_k\})$ with $\sigma = 10^{-4}$ and use it to find noisy versions of $f_\text{VPDR}(\{\tau_k\})$ and $f_\text{R}(\{\tau_k\})$, again reducing the noise in $f_\text{R}(\{\tau_k\})$ by $\sqrt{N}$, where N is the number of MW pulse durations employed for the VPDR signal. Next we fit the noisy signals $f_\text{VPDR}(\{\tau_k\})$ and $f_\text{R}(\{\tau_k\})$ to decaying sinusoids, using a single sinusoid when only $m_I = 0$ is included and using three sinusoids when the full hyperfine triplet is simulated; these fits allow us to extract a (noisy) estimate of the Larmor frequency $\omega_L$. We repeat the Monte Carlo calculation 100 times for each $\tau_\text{max}$ and estimate the axial magnetic field uncertainty $\Delta B^\text{fit}_p$ from the standard deviation in $\omega_L$. Finally, we take the ratio $\Delta B^\text{fit}_p \sqrt{M}/\Delta B^\text{opt}_p$, rescaling the noise to account for the number $M$ of free evolution times employed in the fit.

Unsurprisingly, the sensitivity cost to fitting many free evolution times (rather than measuring only at $\tau_\text{opt}$) is the same for VPDR and DQ Ramsey measurements - as expected, since they have the same functional dependence on $\tau$. The cost is higher when 
hyperfine structure is included, reaching $\sim5\times$ for the free evolution ranges employed in Fig.~\ref{fig4}-\ref{fig6}. 
The free-evolution-range cost to sensitivity shown in Fig.~8b is representative of what we see for different NV orientations and different magnetic fields (except near the dead zones, where the cost increases). To compare the sensitivity of a full 2D spectrum to DQ Ramsey measured at $\tau_\text{opt}$, one would need to multiply the factor from Fig.~8a by the factor from Fig.~8b; in this example, the sensitivity would be roughly 30 times worse. Nevertheless, the factor from Fig.~8a alone is arguably the fairer comparison, since it makes similar assumptions about the measurement context for VPDR and DQ Ramsey. 

Finally, it is worth emphasizing that Eq.'s~\ref{eqn:sensitivityratio}-\ref{eqn:sensitivityratio2} and the results shown in Fig.~8 are specific to the inner product analysis along the $t$ dimension. For example, one might conceivably improve the sensitivity by performing multiple inner products on the same data set to pick out the double quantum signals modulated by both $\Omega$ and $2\Omega$, and extract information from both results. Thus there is room for future optimization of inversion strategies in both the $t$ and $\tau$ dimensions.

\subsection{Sensitivity loss as $\omega_L \rightarrow 0$ \label{app:sensitivity_loss}}
As shown in section~\ref{app:senscomparison}, the signal sensitivity to changes in magnetic field depends on $df/d\omega_L \propto s(\tau) = e^{-2\tau/T_2^*}\tau \sin{2\omega_L\tau}$, where we have now included an exponential dephasing rate as appropriate for typical ensemble samples. As $\omega_L \rightarrow 0$, $s(\tau) \rightarrow 0$ and the protocol becomes insensitive to changes in magnetic field. %

Practically, the reason for loss of sensitivity as $\omega_L \rightarrow 0$ is that we have not allowed dynamic control over the direction of the MW field. To shift the phase of the sinusoidal oscillation in $f_\text{VPDR}(\tau)$ from $\cos{2\omega_L\tau}$ to $\sin{2\omega_L\tau}$ (which would retain sensitivity at vanishing $\omega_L$) requires changing the phase of the bright state superposition $|B\rangle$ targeted by the second microwave pulse relative to the first. While double-quantum Ramsey protocols in a large bias field can spectrally isolate the $|0\rangle \rightarrow|+1\rangle$ and $0\rangle \rightarrow |-1\rangle$ transitions, allowing dynamic control over the bright-state phase~\cite{hart_nv-diamond_2021}, in the absence of a bias field both transitions are targeted by the same MW frequency, and the bright-state phase is determined by the MW polarization.  For example, if we changed the direction of the linearly polarized MW field (written in the NV coordinate frame) from $\hat{x}$ during the first pulse to $\cos{\theta}\hat{x} + \sin{\theta}\hat{y}$ during the second pulse, then Eq.~\ref{eq:P0} would be modified to
\begin{eqnarray}
    P_0(t, \tau) &\approx& \cos^4{\frac{\Omega t}{2}} \nonumber+ \frac{\cos{(\delta \tau + \phi)}}{2}\sin^2{\Omega t}\cos{(\omega_L  \tau - \theta)}\nonumber\\&&+ \sin^4{\frac{\Omega t}{2}}\cos^2{(\omega_L \tau - \theta)}.
    \label{eq:P0theta}
\end{eqnarray}
For $\theta = \pi/4$, we would regain optimal sensitivity at vanishing $\omega_L$. However, given the practical challenge of implementing such a shift in direction while maintaining constant Rabi magnitude for each orientation, we have restricted our calculations to constant MW direction.

We can use the limiting behavior of $s(\tau)$ as $\omega_L \rightarrow 0$ to estimate the size of the associated dead zone. 
While for large $\omega_L \gg 1/T_2^*$ the maximum value of $s$ occurs near $\tau = T_2^*/2$ with a typical magnitude of $T_2^*/2e$, at small $\omega_L \ll 1/T_2^*$ it vanishes linearly with $\omega_L$, i.e. $s(\tau) \approx 2 \tau^2 \omega_L e^{-2\tau/T_2^*}$, with a maximum value $s(T_2^*)\approx 2 (T_2^*)^2 \omega_L/e^2$.  If we require that the magnetometer signal $df/d\omega_L$ stay within a factor of $\epsilon$ of its value for large $\omega_L$, we are restricted to operating with $\omega_L > \epsilon \frac{e}{4 T_2^*}$. For a $T_2^*$ value of $2~\upmu$s, we thus require an axial magnetic field $ \gtrsim 2 \epsilon ~\upmu$T, which is much less than the Earth's magnetic field (25-65 $\upmu$T). For $\epsilon = 0.25$, the $\hat{z}$ component of a $50~\upmu$T magnetic field only violates this bound for angles within $\pm 0.6\degree$ of the x-y plane. 
Numeric simulations also support the conclusion that the angular dead zones are small for magnetic field magnitudes in the range of Earth's field. 

\section{\label{app:numerics} Simulator approximations}

We consider simulated data sets comprising 100's to 1000's of values of $t$ and $\tau$, corresponding to $10^4-10^6$ $(t,\tau)$ pairs, and compute them for many values of the external magnetic field. To permit the simulations to run in reasonable time on standard computing hardware, we make several approximations that greatly speed up execution time. 

First, we neglect the nonsecular terms of the hyperfine interaction with the $^{14}N$ nuclear spin. In the absence of a bias field, the transitions driven by the nonsecular terms are strongly suppressed by $\sim A_\perp/\Delta$, where $A_\perp/(2\pi) =2.7 $~MHz~\cite{felton_hyperfine_2009} and $\Delta/(2\pi) = 2.87$ GHz is the zero-field splitting. While the resulting small shifts in eigenvalues could impact accurate conversion of $\Delta \omega_i$ to magnetic field (especially near level-anti-crossings), such considerations are shared by all magnetometer schemes that detect spin transition frequencies. By neglecting off-axis terms, t
he hyperfine interaction can then be treated as an effective magnetic field $\hbar A m_I /g\mu_B$ along the NV symmetry axis $\hat{z}$ of each orientation, where $A/(2\pi) = 2.16$ MHz~\cite{smeltzer_robust_2009} is the parallel component of the hyperfine tensor. In the secular approximation, the $^{14}$N nuclear spin evolves negligibly during the pulse sequence since Larmor precession about the external field is suppressed by the quadrupolar splitting, so we approximate $m_I$ as a random constant uniformly distributed between \{-1, 0, 1\}. This approximation reduces the dimensionality of the matrix describing the Hamiltonian for each NV orientation from $9\times9$ to $3\times3$.

Next, we approximate the driven evolution in the presence of an off-axis magnetic field by making the rotating wave approximation in the free-Hamiltonian eigenbasis. In the laboratory frame, the Hamiltonian for a given NV orientation and nuclear spin projection $m_I$ is:
\begin{eqnarray}
    \hat{H}_\text{lab} &=& \overbrace{\hbar \Delta \hat{S}_z^2 + \left(g\mu_B\mathbf{B_\text{DC}} + \hbar A~ m_I\hat{z}\right)\cdot \mathbf{\hat{S}}}^{\hat{H}_0}\nonumber\\
    && + g \mu_B \mathbf{B_\text{MW}} \cos{(\nu_{MW}t + \phi)} \cdot \mathbf{\hat{S}},
\end{eqnarray}
where the external DC ($\mathbf{B_\text{DC}}$) and MW ($\mathbf{B_\text{MW}}$) magnetic fields are written in a coordinate system with $\hat{z} ||$ NV symmetry axis and we have indicated the free Hamiltonian terms as $\hat{H}_0$.   
The Schr\"{o}dinger equation (or Master equation) for $\hat{H}_\text{lab}$ can be solved by numerical integration in the laboratory frame, but for $\nu_{MW} \gg \Omega $ it is very inefficient. Instead, we perform the rotating wave approximation as follows: we first diagonalize  $\hat{H}_0$, obtaining the matrix of eigenvectors $\hat{V}$ such that $\hat{V}^\dagger \hat{H}_0 \hat{V} = D$, with $D$ diagonal. We then perform a transformation to a rotating frame in the diagonal basis, 
\begin{equation}
    \hat{H}_\text{rot}^{D} = \hat{U}^\dagger \hat{V}^\dagger \hat{H}_\text{lab} \hat{V} \hat{U} - i \hat{U}^\dagger \frac{d\hat{U}}{dt},
\end{equation}
using a diagonal matrix $\hat{U}$ with entries $e^{-i \nu_{MW}t}$ at positions corresponding to $m_s = \pm1$-like eigenvectors and $1$ at the position corresponding to the $m_s = 0$-like eigenvector of the free Hamiltonian (these assignments are well-defined because we are working in very small magnetic fields $g\mu_B B_{DC} \ll \hbar \Delta$). We then neglect the remaining time-dependent terms in $\hat{H}_\text{rot}^D$. 
This step neglects counter-rotating terms whose effects may become appreciable at the high Rabi frequencies considered, causing Bloch-Siegert shifts of order $\Omega^2/4\Delta \sim $ MHz. We nevertheless drop them because the VPDR protocol is insensitive to detuning.
We thus obtain a 3x3 constant-coefficient matrix $\hat{H}_{RWA}^D$ representing the rotating-wave-approximation Hamiltonian in the free Hamiltonian eigenbasis. 

Finally, we include dephasing within a Lindblad master equation description. While quasi-static noise would be a better approximation of experimental conditions, the convolutions its calculation entails are prohibitively slow; Markovian noise is faster to compute and it produces the same exponential Ramsey decay observed in experiment, 
so we use it to estimate the impacts of dephasing on the protocol. Note, however, that it will not generate the phase shifts observed from quasi-static baths~\cite{koppens_universal_2007}, which is why a cosine inner product is sufficient for inversion of numerical simulations whereas a complex exponential inner product is employed for experimental data analysis. For the simulation, we assume Markovian magnetic noise, corresponding to the jump operators $c_0 = |+1\rangle\langle+1|, c_1 = |-1\rangle\langle-1|$ with equal dephasing rates $\gamma_i = \sqrt{2}/T_2^*$; we represent the jump operators in the free Hamiltonian eigenbasis. Each NV orientation and nuclear spin projection thus obeys
\begin{equation}
\frac{d\rho}{dt} = -\frac{i}{\hbar}\left[\hat{H}_{RWA}, \rho\right] + \sum_i \gamma_i \left(c_i \rho c_i^\dagger - \frac{1}{2}\left\{c_i^\dagger c_i, \rho \right\}\right)
\label{eq:lindblad}
\end{equation}
where $\rho$ is the density matrix. Recasting $\rho$ as a vector, Eq.~\ref{eq:lindblad} can be written as a matrix equation $\frac{d\rho}{dt} = \mathcal{L} \rho$ with $\mathcal{L}$ a $9\times 9$ constant-coefficient matrix. 

By constructing the simulator to solve first-order linear differential equations, we allow fast simulation of experiments performed over an evenly spaced range of $N$ pulse durations $t_j$ and $M$ interpulse delays $\tau_k$. For each pulse duration $t_j = j \Delta t$ with $i = 0,1,2\dots$, we can find the propagator $U_\text{Rabi}(t_j) =(e^{\mathcal{L}\Delta t})^j$ corresponding to evolution under MW driving for duration $t_j$. Since we wish to calculate all of the $U_\text{Rabi}(t_j)$ to simulate all of the pulse durations, we can do it very efficiently by calculating $U_\text{Rabi}(t_1) = e^{\mathcal{L}\Delta t}$ once and then recursively multiplying matrices to find $U_\text{Rabi}(t_{j+1}) = e^{\mathcal{L}\Delta t}U_\text{Rabi}(t_j)$. Similarly, we can recursively find all of the $U_\text{Ramsey}(\tau_k) =(e^{\mathcal{L}_0\Delta \tau})^k$ where $\mathcal{L}_0$ lacks driving terms. For speed, we do these calculations in the diagonalized basis. Once the propagators are calculated, $P_0(t_j,\tau_k)$ can be determined from the appropriate component of $U_\text{Rabi}(t_j)U_\text{Ramsey}(\tau_k)U_\text{Rabi}(t_j)$. 

This calculation is performed for each NV orientation and nuclear spin projection $m_I$; averaging over the 12 signals, equally weighted, yields the ensemble outcome $P_0^\text{total}$. Finally, having computed $P_0^\text{total}$ with $\phi = 0$ we repeat the calculation with $\phi = \pi$ and add the results to obtain the $SQ$-cancelled signal $S_\text{VPDR}$ fed into our data analysis. 


\section{Experimental field reconstruction}
\label{app:experimental}
In this section, we describe how we fit the measured VPDR signals to extract a magnetic field in crystal coordinates that is linearly increasing in coil voltage. From the locations of the inner-product maxima in Fig.~\ref{fig7}c, we obtain $(\omega_\text{max,i}, \nu_\text{max,i})$ for one of the $m_I = +1$ features of each orientation $i$. We estimate the magnitude of the field projection on the $i^\text{th}$ orientation's symmetry axis $|\beta_i^\text{exp}|$ by calculating  
\begin{equation}
|\beta_i^\text{exp}| = |\hbar(\omega_\text{max, i} - 2A)/(2 g \mu_B)|,
\end{equation}
where we subtract off the hyperfine splitting to account for using the $m_i = +1$ feature.
This is an approximation valid for small off-axis fields $\beta_\perp$ relative to the zero-field splitting, with corrections of order $(g \mu_B)\beta_\perp^2/\hbar \Delta$, which even for $\beta_\perp = 100\upmu$T is $\sim 0.1~\upmu$T. We obtain $|\beta_i^\text{exp}(V_n)|$ for the range of coil voltages $V_n$ at which we perform experiments. 

Next, we parameterize the external field in the conventional diamond unit cell coordinates ($\hat{x}\parallel\langle100\rangle, \hat{y}\parallel\langle010\rangle, \hat{z}\parallel\langle001\rangle)$ as $B_a^{(C)}(V) = b_{a} + s_a \times V$, where $a \in \{x,y,z\}$, $(C)$ indicates that we are working in crystal coordinates, $V$ is the coil voltage, and $\{b_x, b_y, bz\}$ and $\{s_x, s_y, s_z\}$ are offset fields and slopes that we seek to determine. The projections of this linear field on each NV axis $\hat{z}_i$ can then be found via
$B_i(V) = \hat{z}_i \cdot \vec{B}^{(C)}(V)$.  Finally we minimize
\begin{equation}
\chi^2  =  \sum_{n,i} \left(|\beta_i^\text{exp}(V_n)| - |B_i(V_n)|\right)^2
\end{equation}
to obtain the offsets $b_a$ and slopes $s_a$ that best fit a linear voltage dependence for the magnetic field. The best-fit offsets and slopes permit us to calculate both $|B_i(V_n)|$ (as shown in Fig.~\ref{fig7}e) and $B_a^{(C)}(V)$ (as shown in Fig.~\ref{fig7}f). Note that this procedure does not have a unique solution; we arbitrarily pick one choice of the possible permutations of NV axes and magnetic field projection signs by virtue of the initial guesses that we provide to the fit. Nevertheless, even with this degeneracy, the fact that we can find a good fit to all four orientations' signals with three vector field components demonstrates that the VPDR data are mutually consistent.
\input{main.bbl}
\end{document}

%% file: main.bbl
%